# Sparsity-Cognizant Total Least-Squares for Perturbed Compressive Sampling[†]

*Hao Zhu*[*], *Geert Leus*[⋆]*, and Georgios B. Giannakis*[*] *(contact author)*

**Submitted:** May 30, 2018

## Abstract

Solving linear regression problems based on the total least-squares (TLS) criterion has well-documented merits in various applications, where perturbations appear both in the data vector as well as in the regression matrix. However, existing TLS approaches do not account for sparsity possibly present in the unknown vector of regression coefficients. On the other hand, sparsity is the key attribute exploited by modern compressive sampling and variable selection approaches to linear regression, which include noise in the data, but do not account for perturbations in the regression matrix. The present paper fills this gap by formulating and solving TLS optimization problems under sparsity constraints. Near-optimum and reduced-complexity suboptimum sparse (S-) TLS algorithms are developed to address the perturbed compressive sampling (and the related dictionary learning) challenge, when there is a mismatch between the true and adopted bases over which the unknown vector is sparse. The novel S-TLS schemes also allow for perturbations in the regression matrix of the least-absolute selection and shrinkage selection operator (Lasso), and endow TLS approaches with ability to cope with sparse, under-determined "errors-in-variables" models. Interesting generalizations can further exploit prior knowledge on the perturbations to obtain novel weighted and structured S-TLS solvers. Analysis and simulations demonstrate the practical impact of S-TLS in calibrating the mismatch effects of contemporary grid-based approaches to cognitive radio sensing, and robust direction-of-arrival estimation using antenna arrays.

**EDICS:** SSP-SNMD, SSP-PARE, DSP-RECO, SAM-CALB, MLR-COGP

[†] This work is supported by the NSF grants CCF-0830480, CCF-1016605, ECCS-0824007, and ECCS-1002180; G. Leus is supported in part by NWO-STW under the VICI program (project 10382). Part of this work was presented at the *11th IEEE International Workshop on Signal Processing Advances in Wireless Communications*, Marrakech, Marocco, June 20 – 23, 2010.

[*]H. Zhu and G. B. Giannakis are with the Department of Electrical and Computer Engineering, University of Minnesota, 200 Union Street SE, Minneapolis, MN 55455, USA. Tel/fax: (612)624-9510/625-2002, emails: {zhuh,georgios}@umn.edu

[⋆]G. Leus is with the Faculty of Electrical Engineering, Mathematics and Computer Science, Delft University of Technology, Mekelweg 4, 2628 CD Delft, The Netherlands. Tel/fax: +31-15-2784327/2786190, email: g.j.t.leus@tudelft.nl

# I. INTRODUCTION

Sparsity is an attribute possessed by many signal vectors either naturally, or, after projecting them over appropriate bases. It has been exploited for a while in numerical linear algebra, statistics, and signal processing, but renewed interest emerged in recent years because sparsity plays an instrumental role in modern compressive sampling (CS) theory and applications; see e.g., [3].

In the noise-free setup, CS holds promise to address problems as fundamental as solving exactly under-determined systems of linear equations when the unknown vector is sparse [8]. Variants of CS for the "noisy setup" are rooted in the basis pursuit (BP) approach [11], which deals with fitting sparse linear representations to perturbed measurements – a task of major importance for signal compression and feature extraction. The Lagrangian form of BP is also popular in statistics for fitting sparse linear regression models, using the so-termed least-absolute shrinkage and selection operator (Lasso); see e.g., [28], [19], and references thereof. However, existing CS, BP, and Lasso-based approaches do not account for perturbations present in the matrix of equations, which in the BP (respectively Lasso) parlance is referred to as the representation basis or dictionary (correspondingly regression) matrix.

Such perturbations appear when there is a *mismatch* between the adopted basis matrix and the actual but *unknown* one – a performance-critical issue in e.g., sparsity-exploiting approaches to localization, time delay, and Doppler estimation in communications, radar, and sonar applications [16], [22], [6], [2]. Performance analysis of CS and BP approaches for the partially-perturbed linear model with perturbations only in the basis matrix, as well as for the fully-perturbed one with perturbations present also in the measurements, was pursued recently in [20], [12], and [10]. But devising a systematic approach to reconstructing sparse vectors under either type of perturbed models was left open.

Interestingly, for *non-sparse* over-determined linear systems, such an approach is available within the framework of total least-squares (TLS), the basic generalization of LS tailored for fitting fully-perturbed linear models [30]. TLS and its variants involving regularization with the $\ell_2$-norm of the unknown vector [26], have found widespread applications in diverse areas, including system identification with errors-in-variables (EIV), retrieval of spatial and temporal harmonics, reconstruction of medical images, and forecasting of financial data [23]. TLS was also utilized by [13] for dictionary learning, but the problem reduces to an over-determined linear system with a non-sparse unknown vector. Unfortunately, TLS approaches, with or without existing regularization terms, cannot yield consistent estimators when the linear model is under-determined, nor they account for sparsity present in the unknown vector of regression coefficients.

From a high-level vantage point, the present paper is about fitting sparse, perturbed, linear models, through what is termed here the sparse (S-) TLS framework. On the one hand, S-TLS provides CS, BP, and Lasso-type algorithms suitable for fitting partially- and fully-perturbed linear models. On the other hand, it furnishes TLS with sparsity-cognizant regularized alternatives, which yield consistent estimators even for under-determined models. The novel framework does not require a priori information on the underlying perturbations, and in this sense S-TLS based algorithms have universal applicability. However, the framework is flexible enough to accommodate both deterministic as well as probabilistic prior information that maybe available on the perturbed data. The practical impact is apparent to any CS, BP/Lasso, and TLS-related application involving reconstruction of a sparse vector based on data adhering to an over- or under-determined, partially- or fully-perturbed, linear model.

The specific contributions and organization of the paper are as follows. With unifying notation, Section II outlines the pertinent CS-BP-Lasso-TLS context, and introduces the S-TLS formulation and problem statement. Section III presents two equivalent formulations, which are first used to establish optimality of S-TLS estimators in the maximum a posteriori (MAP) sense, under a fully-perturbed EIV model. Subsequently, the same formulations are utilized in Section IV to develop near-optimum and reduced-complexity suboptimum S-TLS solvers with convergence guarantees. The scope of S-TLS is considerably broadened in Section V, where a priori information on the deterministic structure of the data vector, the basis matrix, and/or the statistics of the perturbations is incorporated to develop weighted and structured (WS) S-TLS criteria along with associated algorithms to optimize them. The impact of WSS-TLS is demonstrated in Section VI using two paradigms: cognitive radio sensing, and direction of arrival estimation with (possibly uncalibrated) antenna arrays. Simulated tests in Section VII illustrate the merits of the novel (WS)S-TLS framework relative to BP, Lasso, and TLS alternatives. The paper is wrapped up with brief concluding comments and future research directions in Section VIII.

*Notation:* Upper (lower) bold face letters are used throughout to denote matrices (column vectors); $(\cdot)^T$ denotes transposition; $(\cdot)^\dagger$ the matrix pseudo-inverse; $\text{vec}(\cdot)$ the column-wise matrix vectorization; $\otimes$ the Kronecker product; $\lceil \cdot \rceil$ the ceiling function; $\mathbf{1}_{m \times n}$ the $m \times n$ matrix of all ones; $\mathbf{0}_{m \times n}$ the $m \times n$ matrix of all zeros; $\mathbf{I}$ the identity matrix; $\|\cdot\|_F$ the Frobenius norm; and $\|\cdot\|_p$ the $p$-th vector norm for $p \geq 1$; $\mathcal{N}(\boldsymbol{\mu}, \boldsymbol{\Sigma})$ the vector Gaussian distribution with mean $\boldsymbol{\mu}$ and covariance $\boldsymbol{\Sigma}$; and $p[x = x'|y = y']$ the conditional probability density function (pdf) of the continuous random variable (r.v.) $x$ taking the value $x'$, given that the r.v. $y$ took the value $y'$.

## II. Preliminaries and Problem Statement

Consider the *under-determined* linear system of equations, $\mathbf{y} = \mathbf{C}\boldsymbol{\theta}_o$, where the unknown $n \times 1$ vector $\boldsymbol{\theta}_o$ is to be recovered from the given $m \times 1$ data vector $\mathbf{y}$ and the $m \times n$ matrix $\mathbf{C}$. With $m < n$ and no further assumption, only approximations of $\boldsymbol{\theta}_o$ are possible using the minimum-norm solution; or, the least-squares (LS) regularized by the $\ell_2$-norm, which solves in closed form the quadratic problem: $\min_{\boldsymbol{\theta}} \|\mathbf{y} - \mathbf{C}\boldsymbol{\theta}\|_2^2 + \gamma\|\boldsymbol{\theta}\|_2^2$ for some chosen $\gamma > 0$. Suppose instead that over a known basis matrix $\mathbf{B}$, the unknown vector satisfies $\boldsymbol{\theta}_o = \mathbf{B}\mathbf{x}_o$ with $\mathbf{x}_o$ being *sparse*, meaning that:

(**as0**) *The $n \times 1$ vector $\mathbf{x}_o$ contains more than $n - m$ zero elements at unknown entries.*

Under (as0) and certain conditions on the matrix $\mathbf{A} := \mathbf{C}\mathbf{B}$, compressive sampling (CS) theory asserts that exact recovery of $\mathbf{x}_o$ can be guaranteed by solving the nonconvex, combinatorially complex problem: $\min_{\mathbf{x}} \|\mathbf{x}\|_0$ subject to (s.to) $\mathbf{y} = \mathbf{A}\mathbf{x}$. More interestingly, the same assertion holds with quantifiable chances if one relaxes the $\ell_0$- via the $\ell_1$-norm, and solves efficiently the convex problem: $\min_{\mathbf{x}} \|\mathbf{x}\|_1$ s.to $\mathbf{y} = \mathbf{A}\mathbf{x}$ [3], [8], [11].

Suppose now that due to data perturbations the available vector $\mathbf{y}$ adheres only approximately to the linear model $\mathbf{A}\mathbf{x}_o$. The $\ell_1$-norm based formulation accounting for the said perturbations is known as basis pursuit (BP) [11], and the corresponding convex problem written in its Lagrangian form is: $\min_{\mathbf{x}} \|\mathbf{y} - \mathbf{A}\mathbf{x}\|_2^2 + \lambda_1 \|\mathbf{x}\|_1$, where $\lambda_1 > 0$ is a sparsity-tuning parameter. (For large $\lambda_1$, the solution is driven toward the all-zero vector; whereas for small $\lambda_1$ it tends to the LS solution.) This form of BP coincides with the Lasso approach developed for variable selection in linear regression problems [19], [28]. For uniformity with related problems, the BP/Lasso solvers can be equivalently written as

$$\{\hat{\mathbf{x}}_{Lasso}, \hat{\mathbf{e}}_{Lasso}\} := \arg\min_{\mathbf{x}, \mathbf{e}} \ \|\mathbf{e}\|_2^2 + \lambda_1 \|\mathbf{x}\|_1 \tag{1a}$$

$$\text{s. to} \ \ \mathbf{y} + \mathbf{e} = \mathbf{A}\mathbf{x}. \tag{1b}$$

Two interesting questions arise at this point: i) How is the performance of CS and BP/Lasso based reconstruction affected if perturbations appear also in $\mathbf{A}$? and ii) How can sparse vectors be efficiently reconstructed from over- and especially under-determined linear regression models while accounting for perturbations present in $\mathbf{y}$ and/or $\mathbf{A}$?

In the context of CS, perturbations in $\mathbf{A}$ can be due to disturbances in the compressing matrix $\mathbf{C}$, in the basis matrix $\mathbf{B}$, or in both. Those in $\mathbf{C}$ can be due to non-idealities in the analog implementation of CS; while those in $\mathbf{B}$ can also emerge because of mismatch between the adopted basis $\mathbf{B}$ and the actual one, which being unknown, is modeled as $\mathbf{B} + \mathbf{E}_B$. This mismatch emerges with grid-based approaches to localization, time delay, and spatio-temporal frequency or Doppler estimation [2], [4], [6], [9], [16], [17]. In these applications, the entries of $\boldsymbol{\theta}_o$ have e.g., a sparse discrete-time Fourier transform with peaks off the frequency grid $\{2\pi k/n\}_{k=0}^{n-1}$, but the postulated $\mathbf{B}$ is the fast Fourier transform (FFT) matrix built from this canonical grid. In this case, the actual linear relationship is $\boldsymbol{\theta}_o = (\mathbf{B} + \mathbf{E}_B)\mathbf{x}_o$ with $\mathbf{x}_o$ sparse. Bounds on the CS reconstruction error under basis mismatch are provided in [12]; see also [10], where the mismatch-induced error was reduced by increasing the grid density. Performance of BP/Lasso approaches for the under-determined, fully-perturbed (in both $\mathbf{y}$ and $\mathbf{A}$) linear model was analyzed in [20] by bounding the reconstruction error, and comparing it against its counterpart derived for the partially-perturbed (only in $\mathbf{y}$) model derived in [8]. Collectively, [12] and [20] address the performance question i), but provide no algorithms to address the open research issue ii).

The overarching theme of the present paper is to address this issue by developing a sparse total least-squares (S-TLS) framework. Without exploiting sparsity, TLS has well-documented impact in applications as broad as linear prediction, system identification with errors-in-variables (EIV), spectral analysis, image reconstruction, speech, and audio processing, to name a few; see [30] and references therein. For over-determined models with unknown vectors $\mathbf{x}_o$ not abiding with (as0), TLS estimates are given by

$$\left\{\hat{\mathbf{x}}_{TLS}, \hat{\mathbf{E}}_{TLS}, \hat{\mathbf{e}}_{TLS}\right\} := \arg\min_{\mathbf{x},\mathbf{E},\mathbf{e}} \ \|[\mathbf{E}\ \mathbf{e}]\|_F^2 \tag{2a}$$

$$\text{s. to}\ \ \mathbf{y} + \mathbf{e} = (\mathbf{A} + \mathbf{E})\mathbf{x}. \tag{2b}$$

To cope with ill-conditioned matrices $\mathbf{A}$, an extra constraint bounding $\|\boldsymbol{\Gamma}\mathbf{x}\|_2$ is typically added in (2) to obtain different regularized TLS estimates depending on the choice of matrix $\boldsymbol{\Gamma}$ [5], [26].

The distinct objective of S-TLS relative to (regularized) TLS is twofold: account for sparsity as per (as0), and develop S-TLS solvers especially for under-determined, fully-perturbed linear models. To accomplish these goals, one must solve the S-TLS problem concretely formulated (for $\lambda > 0$) as [cf. (1), (2)]

$$\left\{\hat{\mathbf{x}}_{S-TLS}, \hat{\mathbf{E}}_{S-TLS}, \hat{\mathbf{e}}_{S-TLS}\right\} := \arg\min_{\mathbf{x},\mathbf{e},\mathbf{E}} \ \|[\mathbf{E}\ \mathbf{e}]\|_F^2 + \lambda\|\mathbf{x}\|_1 \tag{3a}$$

$$\text{s. to}\ \ \mathbf{y} + \mathbf{e} = (\mathbf{A} + \mathbf{E})\mathbf{x}\,. \tag{3b}$$

The main goal is to develop efficient algorithms attaining at least the local and hopefully the global optimum of (3) – a challenging task since presence of the product $\mathbf{E}\mathbf{x}$ reveals that the problem is generally nonconvex. Similar to LS, BP, Lasso, and TLS, it is also worth stressing that the S-TLS estimates sought in (3) are universal in the sense that perturbations in $\mathbf{y}$ and $\mathbf{A}$ can be random or deterministic with or without a priori known structure.

But if prior knowledge is available on the perturbations, can weighted and structured S-TLS problems be formulated and solved? Can the scope of S-TLS be generalized (e.g., to recover a sparse matrix $\mathbf{X}_o$ using $\mathbf{A}$ and a data matrix $\mathbf{Y}$), and thus have impact in classical applications such as calibration of antenna arrays, or contemporary ones, such as cognitive radio sensing? Can S-TLS estimates be (e.g., Bayes) optimal if additional modeling assumptions are invoked? These questions will be addressed in the ensuing sections, starting from the last one.

### III. MAP Optimality of S-TLS for EIV Models

Consider the EIV model with perturbed input ($\mathbf{A}$) and perturbed output ($\mathbf{y}$) obeying the relationship

$$\mathbf{y} = \mathbf{A}_o \mathbf{x}_o + (-\mathbf{e}_y), \quad \mathbf{A} = \mathbf{A}_o + (-\mathbf{E}_A) \qquad (4)$$

where the notation of the model perturbations $\mathbf{e}_y$ and $\mathbf{E}_A$ stresses their difference with $\mathbf{e}$ and $\mathbf{E}$, which are variables selected to yield the optimal S-TLS fit in (3). In a system identification setting, $\mathbf{e}_y$ and $\mathbf{E}_A$ are stationary random processes giving rise to noisy output/input data $\mathbf{y}/\mathbf{A}$, based on which the task is to estimate the system vector $\mathbf{x}_o$ (comprising e.g., impulse response or pole-zero parameters), and possibly the inaccessible input matrix $\mathbf{A}_o$. To assess statistical optimality of the resultant estimators, collect the model perturbations in a column-vector form as $\text{vec}([\mathbf{E}_A \; \mathbf{e}_y])$, and further assume that:

(**as1**) *Perturbations of the EIV model in* (4) *are independent identically distributed (i.i.d.), Gaussian r.v.s, i.e., $\text{vec}([\mathbf{E}_A \; \mathbf{e}_y]) \sim \mathcal{N}(\mathbf{0}, \mathbf{I})$, independent from $\mathbf{A}_o$ and $\mathbf{x}_o$. Entries of $\mathbf{x}_o$ are zero-mean, i.i.d., according to a common Laplace distribution. In addition, either* (**a**) *the entries of $\mathbf{x}_o$ have common Laplacian parameter $2/\lambda$, and are independent from $\mathbf{A}_o$, which has i.i.d. entries drawn from a zero-mean uniform (i.e., non-informative) prior pdf; or,* (**b**) *the common Laplacian parameter of $\mathbf{x}_o$ entries is $2(\sigma^2+1)/(\lambda\sigma^2)$, and $\mathbf{A}_o$ conditioned on $\mathbf{x}_o$ has i.i.d. rows with pdf $\mathcal{N}(\mathbf{0}, \sigma^2[\mathbf{I} - (1 + \|\mathbf{x}_o\|_2^2)^{-1}\mathbf{x}_o\mathbf{x}_o^T])$.*

Note that the heavy-tailed Laplacian prior on $\mathbf{x}_o$ under (as1) is in par with the "non-probabilistic" sparsity attribute in (as0). It has been used to establish that the Lasso estimator in (1) is optimal, in the maximum a posteriori (MAP) sense, when $\mathbf{E}_A \equiv \mathbf{0}$ [28]. If on the other hand, $\mathbf{x}_o$ is viewed as non-sparse, deterministic and $\mathbf{A}_o$ as deterministic or as adhering to (as1b), it is known that the TLS estimator in (2) is optimum in the maximum likelihood (ML) sense for the EIV model in (4); see [23] and [24].

Aiming to establish optimality of S-TLS under (as1), it is useful to re-cast (3) as described in the following lemma. (This lemma will be used also in developing S-TLS solvers in Section IV.)

**Lemma 1:** *The constrained S-TLS formulation in* (3) *is equivalent to two unconstrained (also nonconvex)*

*optimization problems: (a) one involving* $\mathbf{x}$ *and* $\mathbf{E}$ *variables, namely*

$$\left\{\hat{\mathbf{x}}_{S-TLS}, \hat{\mathbf{E}}_{S-TLS}\right\} = \arg\min_{\mathbf{x}, \mathbf{E}} \ \left[\|\mathbf{y} - (\mathbf{A}+\mathbf{E})\mathbf{x}\|_2^2 + \|\mathbf{E}\|_F^2 + \lambda\|\mathbf{x}\|_1\right] \quad (5)$$

*and (b) one of fractional form involving only the variable* $\mathbf{x}$*, expressed as*

$$\hat{\mathbf{x}}_{S-TLS} := \arg\min_{\mathbf{x}} \frac{\|\mathbf{y}-\mathbf{A}\mathbf{x}\|_2^2}{1+\|\mathbf{x}\|_2^2} + \lambda\|\mathbf{x}\|_1. \quad (6)$$

*Proof:* To establish the equivalence of (5) with (3), simply eliminate $\mathbf{e}$ by substituting the constraint (3b) into the cost function of (3a). For (6), let $\mathbf{v} := \text{vec}([\mathbf{E} \ \mathbf{e}])$, and re-write the cost in (3a) as $\|[\mathbf{E} \ \mathbf{e}]\|_F^2 = \|\mathbf{v}\|_2^2$; and the constraint (3b) as $\mathbf{y} - \mathbf{A}\mathbf{x} = \mathbf{G}(\mathbf{x})\mathbf{v}$, where $\mathbf{G}(\mathbf{x}) := \mathbf{I} \otimes [\mathbf{x}^T, -1]$. With $\mathbf{x}$ fixed, the $\ell_1$-norm can be dropped from (3a), and the reformulated optimization becomes: $\min_{\mathbf{v}} \|\mathbf{v}\|_2^2$ s. to $\mathbf{y} - \mathbf{A}\mathbf{x} = \mathbf{G}(\mathbf{x})\mathbf{v}$. But the latter is a minimum-norm LS problem, admitting the closed-form solution

$$\mathbf{v}(\mathbf{x}) = \mathbf{G}^T(\mathbf{x})[\mathbf{G}(\mathbf{x})\mathbf{G}^T(\mathbf{x})]^{-1}(\mathbf{y}-\mathbf{A}\mathbf{x}) = (1+\|\mathbf{x}\|_2^2)^{-1}\mathbf{G}^T(\mathbf{x})(\mathbf{y}-\mathbf{A}\mathbf{x}) \quad (7)$$

where the second equality holds because $\mathbf{G}(\mathbf{x})\mathbf{G}^T(\mathbf{x}) = \|[\mathbf{x}^T, -1]\|_2^2 \mathbf{I} = (1+\|\mathbf{x}\|_2^2)\mathbf{I}$. Substituting (7) back into the cost $\|\mathbf{v}\|_2^2$, yields readily the fractional form in (6), which depends solely on $\mathbf{x}$. ∎

Using Lemma 1, it is possible to establish MAP optimality of the S-TLS estimator as follows.

**Proposition 1:** *(MAP optimality). Under (as1), the S-TLS estimator in* (3) *is MAP optimal for the EIV model in* (4)*. Specifically,* (5) *is MAP optimal for estimating both* $\mathbf{x}_o$ *and* $\mathbf{A}_o$ *under (as1a), while* (6) *is MAP optimal for estimating only* $\mathbf{x}_o$ *under (as1b).*

*Proof:* Given $\mathbf{y}$ and $\mathbf{A}$, the MAP approach to estimating both $\mathbf{x}_o$ and $\mathbf{A}_o$ in (4) amounts to maximizing with respect to (wrt) $\mathbf{x}$ and $\mathbf{E}$ the logarithm of the posterior pdf denoted as $\ln p[\mathbf{x}_o = \mathbf{x}, \mathbf{A}_o = \mathbf{A}+\mathbf{E}|\mathbf{y},\mathbf{A}]$. Recalling that $\mathbf{x}_o$ and $\mathbf{A}_o$ are independent under (as1a), Bayes' rule implies that this is equivalent to: $\min_{\mathbf{x},\mathbf{E}} -\{\ln p[\mathbf{y},\mathbf{A}|\mathbf{x}_o=\mathbf{x},\mathbf{A}_o=\mathbf{A}+\mathbf{E}] + \ln p[\mathbf{x}_o=\mathbf{x}] + \ln p[\mathbf{A}_o=\mathbf{A}+\mathbf{E}]\}$, where the summands correspond to the (conditional) log-likelihood and the log-prior pdfs, respectively. The log-prior associated with the Laplacian pdf of $\mathbf{x}_o$ is given by

$$\ln p[\mathbf{x}_o = \mathbf{x}] = \ln \prod_{\nu=1}^{n}[(\lambda/4)\exp(-\lambda|x_\nu|/2)] = -(\lambda/2)\sum_{\nu=1}^{n}|x_\nu| + n\ln(\lambda/4) \quad (8)$$

while the log-prior associated with the uniform pdf of $\mathbf{A}_o$ is constant under (as1a), and thus does not affect the MAP criterion. Conditioning the log-likelihood on $\mathbf{x}_o$ and $\mathbf{A}_o$, implies that the only sources of randomness in the data $[\mathbf{y} \ \mathbf{A}]$ are the EIV model perturbations, which under (as1) are independent, standardized Gaussian; thus, the conditional log-likelihood is $\ln p[\mathbf{y},\mathbf{A}|\mathbf{x}_o=\mathbf{x},\mathbf{A}_o=\mathbf{A}+\mathbf{E}] = \ln[\mathbf{e}_y = \mathbf{y}-(\mathbf{A}+\mathbf{E})\mathbf{x}] + \ln p[\mathbf{E}_A = \mathbf{E}]$. After omitting terms not dependent on the variables $\mathbf{x}$ and $\mathbf{E}$, the latter shows that the log-likelihood contributes to the MAP criterion two quadratic terms (sum of two Gaussian exponents): $(1/2)\{\|\mathbf{y} - (\mathbf{A}+\mathbf{E})\mathbf{x}\|_2^2 + \|\mathbf{E}\|_F^2\}$. Upon combining these quadratic terms with the $\ell_1$-norm coming from the sum in (8), the log-posterior pdf boils down to the form minimized in (5), which per Lemma 1 is equivalent to (3), and thus establishes MAP optimality of S-TLS under (as1a).

Proceeding to prove optimality under (as1b), given again the data $\mathbf{y}$ and $\mathbf{A}$, consider the MAP approach now to estimate only $\mathbf{x}_o$ in (4), treating $\mathbf{A}_o$ as a nuisance parameter matrix that satisfies (as1b). MAP here amounts to maximizing (wrt $\mathbf{x}$ only) the criterion $\ln p[\mathbf{x}_o = \mathbf{x}|\mathbf{y}, \mathbf{A}]$; and Bayes' rule leads to the equivalent problem $\min_\mathbf{x} -\{\ln p[\mathbf{y}, \mathbf{A}|\mathbf{x}_o = \mathbf{x}] + \ln p[\mathbf{x}_o = \mathbf{x}]\}$. But conditioned on $\mathbf{x}_o$, (as1b) dictates that $\mathbf{A}_o$ and $[\mathbf{E}_A\ \mathbf{e}_y]$ are zero-mean Gaussian and independent. Thus, linearity of the EIV model (4) implies that $\mathbf{y}$ and $\mathbf{A}$ are zero-mean jointly Gaussian in the conditional log-likelihood. Since rows of $\mathbf{A}_o$ and $[\mathbf{E}_A\ \mathbf{e}_y]$ are (conditionally) i.i.d. under (as1b), the rows of matrix $[\mathbf{A}\ \mathbf{y}]$ are independent. In addition, the $\rho$th-row of $[\mathbf{A}\ \mathbf{y}]$ denoted as $[\mathbf{a}_\rho^T\ y_\rho]$, has inverse (conditional) covariance matrix

$$\mathbb{E}\left[\begin{bmatrix}\mathbf{a}_\rho\\y_\rho\end{bmatrix}[\mathbf{a}_\rho^T\ y_\rho]\bigg|\mathbf{x}_o=\mathbf{x}\right]^{-1} = \begin{bmatrix}(\sigma^2+1)\mathbf{I} - \sigma^2\mathbf{x}\mathbf{x}^T/(1+\|\mathbf{x}\|_2^2) & \sigma^2\mathbf{x}/(1+\|\mathbf{x}\|_2^2)\\ \sigma^2\mathbf{x}^T/(1+\|\mathbf{x}\|_2^2) & 1+\sigma^2\|\mathbf{x}\|_2^2/(1+\|\mathbf{x}\|_2^2)\end{bmatrix}^{-1}$$
$$= \frac{1}{\sigma^2+1}\left\{\mathbf{I} + \frac{\sigma^2}{1+\|\mathbf{x}\|_2^2}\begin{bmatrix}\mathbf{x}\\-1\end{bmatrix}[\mathbf{x}^T\ -1]\right\} \quad (9)$$

with determinant $1/(\sigma^2+1)^n$ not a function of $\mathbf{x}$. After omitting such terms not dependent on $\mathbf{x}$, and using the independence among rows and their inverse covariance in (9), the conditional log-likelihood boils down to the fractional form $\frac{\sigma^2}{2(\sigma^2+1)}\|\mathbf{y}-\mathbf{A}\mathbf{x}\|_2^2/(1+\|\mathbf{x}\|_2^2)$. Since the Laplacian parameter under (as1b) equals $2(\sigma^2+1)/(\lambda\sigma^2)$, the log-prior in (8) changes accordingly; and together with the fractional form of the log-likelihood reduces the negative log-posterior to the cost in (6). This establishes MAP optimality of the equivalent S-TLS in (3) for estimating only $\mathbf{x}_o$ in (4), under (as1b). ∎

Proposition 1 will be generalized in Section V to account for structured and correlated perturbations with known covariance matrix. But before pursuing these generalizations, S-TLS solvers of the problem in (3) are in order.

## IV. S-TLS SOLVERS

Two iterative algorithms are developed in this section to solve the S-TLS problem in (3), which was equivalently re-formulated as in (5) and (6). The first algorithm can approach the *global optimum* but is computationally demanding; while the second one guarantees convergence to a *local optimum* but is computationally efficient. Thus, in addition to being attractive on its own, the second algorithm can serve as initialization to speed up convergence (and thus reduce computational burden) of the first one. To appreciate the challenge and the associated performance-complexity tradeoffs in developing algorithms for optimizing S-TLS criteria, it is useful to recall that all S-TLS problems are nonconvex; hence, unlike ordinary TLS that can be globally optimized (e.g., via SVD [23]), no efficient convex optimization solver is available with guaranteed convergence to the global optimum of (3), (5), or (6).

## A. Bisection-based $\varepsilon$-Optimal Algorithm

Viewing the cost in (6) as a Lagrangian function, allows casting this unconstrained minimization problem as a constrained one. Indeed, sufficiency of the Lagrange multiplier theory implies that [7, Sec 3.3.4]: using the solution $\hat{\mathbf{x}}_{S-TLS}$ of (6) for a given multiplier $\lambda > 0$ and letting $\mu := \|\hat{\mathbf{x}}_{S-TLS}\|_1$, the pertinent constraint is $\mathcal{X}_1(\mu) := \{\mathbf{x} \in \mathbb{R}^n : \|\mathbf{x}\|_1 \leq \mu\}$; and the equivalent constrained minimization problem is [cf. (6)]

$$\hat{\mathbf{x}}_{S-TLS} := \arg \min_{\mathbf{x} \in \mathcal{X}_1(\mu)} f(\mathbf{x}), \quad f(\mathbf{x}) := \frac{\|\mathbf{y} - \mathbf{A}\mathbf{x}\|_2^2}{1 + \|\mathbf{x}\|_2^2}. \tag{10}$$

There is no need to solve (6) in order to specify $\mu$, because a cross-validation scheme can be implemented to specify $\mu$ in the stand-alone problem (10), along the lines used by e.g., [25] to determine $\lambda$ in (6). The remainder of this subsection will thus develop an iterative scheme converging to the global optimum of (10), bearing in mind that this equivalently solves (6), (5) and (3) too.

From a high-level view, the novel scheme comprises an *outer iteration* loop based on the bisection method [14], and an *inner iteration* loop that relies on a variant of the branch-and-bound (BB) method [1]. A related approach was pursued in [5] to solve the clairvoyant TLS problem (2) under $\ell_2$-norm regularization constraints. The challenging difference with the S-TLS here is precisely the non-differentiable $\ell_1$-norm constraint in $\mathcal{X}_1(\mu)$. The outer iteration "squeezes" the minimum cost $f(\mathbf{x})$ in (10) between successively shrinking lower and upper bounds expressible through a parameter $a$. Per outer iteration, these bounds are obtained via inner iterations equivalently minimizing a surrogate quadratic function $g(\mathbf{x}, a)$, which does not have fractional form, and is thus more convenient to optimize than $f(\mathbf{x})$.

Given an upper bound $a$ on $f(\mathbf{x})$, the link between $f(\mathbf{x})$ and $g(\mathbf{x}, a)$ follows if ones notes that

$$0 \leq a^\star := \min_{\mathbf{x} \in \mathcal{X}_1(\mu)} f(\mathbf{x}) = \min_{\mathbf{x} \in \mathcal{X}_1(\mu)} \frac{\|\mathbf{y} - \mathbf{A}\mathbf{x}\|_2^2}{1 + \|\mathbf{x}\|_2^2} \leq a \tag{11}$$

is equivalent to

$$g^\star(a) := \min_{\mathbf{x} \in \mathcal{X}_1(\mu)} g(\mathbf{x}, a) = \min_{\mathbf{x} \in \mathcal{X}_1(\mu)} \left\{ \|\mathbf{y} - \mathbf{A}\mathbf{x}\|_2^2 - a(1 + \|\mathbf{x}\|_2^2) \right\} \leq 0. \tag{12}$$

Suppose that after outer iteration $i$ the optimum $a^\star$ in (11) belongs to a known interval $\mathcal{I}_i := [l_i, u_i]$. Suppose further that the inner loop yields the global optimum in (12) for $a = (l_i + u_i)/2$, and consider evaluating the sign of $g^\star(a)$ at this middle point $a = (l_i + u_i)/2$ of the interval $\mathcal{I}_i$. If $g^\star((l_i+u_i)/2) > 0$, the equivalence between (12) and (11) implies that $a^\star > (l_i + u_i)/2 > l_i$; and hence, $a^\star \in \mathcal{I}_{i+1} := [(l_i + u_i)/2, u_i]$, which yields a reduced-size interval $\mathcal{I}_{i+1}$ by shrinking $\mathcal{I}_i$ from the left. On the other hand, if $g^\star((l_i + u_i)/2) < 0$, the said equivalence will imply that $a^\star \in \mathcal{I}_{i+1} := [l_i, (l_i + u_i)/2]$, which shrinks the $\mathcal{I}_i$ interval from the right. This successive shrinkage through bisection explains how the outer iteration converges to the global optimum of (10).

What is left before asserting rigorously this convergence, is to develop the inner iteration which ensures that the global optimum in (12) can be approached for any given $a$ specified by the outer bisection-based iteration. To appreciate the difficulty here note that the Hessian of $g(\mathbf{x}, a)$ is given by $\mathbf{H} := 2(\mathbf{A}^T\mathbf{A} - a\mathbf{I})$. Clearly, $\mathbf{H}$ is not guaranteed to be positive or negative definite since $a$ is positive. As a result, the cost $g(\mathbf{x}, a)$ in (12) bypasses the fractional form of $f(\mathbf{x})$ but it is still an indefinite quadratic, and hence nonconvex. Nonetheless, the quadratic form of $g(\mathbf{x}, a)$ allows adapting the BB iteration of [1], which can yield a feasible and $\delta$-optimum solution $\mathbf{x}_g^\star$ satisfying: a) $\mathbf{x}_g^\star \in \mathcal{X}_1(\mu)$; and b) $g^\star(a) \le g(\mathbf{x}_g^\star, a) \le g^\star(a) + \delta$, where $\delta$ denotes a pre-specified margin.

In the present context, the BB algorithm finds successive upper and lower bounds of the function

$$g_{\text{box}}^\star(a) := \min_{\mathbf{x} \in \mathcal{X}_1(\mu), \mathbf{x}_L \le \mathbf{x} \le \mathbf{x}_U} g(\mathbf{x}, a) \tag{13}$$

where the constraint $\mathbf{x}_L \le \mathbf{x} \le \mathbf{x}_U$ represents a box that shrinks as iterations progress. Upon converting the constraints of (13) to linear ones, upper bounds $\mathcal{U}$ on the function $g_{\text{box}}^\star(a)$ in (13) can be readily obtained via suboptimum solvers of the constrained optimization of the indefinite quadratic cost $g(\mathbf{x}, a)$; see e.g., [7, Chp. 2]. Lower bounds on $g_{\text{box}}^\star(a)$ can be obtained by minimizing a convex function $g_L(\mathbf{x}, a)$, which under-approximates $g(\mathbf{x}, a)$ over the interval $\mathbf{x}_L \le \mathbf{x} \le \mathbf{x}_U$. This convex approximant is given by

$$g_L(\mathbf{x}, a) = g(\mathbf{x}, a) + (\mathbf{x} - \mathbf{x}_L)^T \mathbf{D}(\mathbf{x} - \mathbf{x}_U) \tag{14}$$

where $\mathbf{D}$ is a diagonal positive semi-definite matrix chosen to ensure that $g_L(\mathbf{x}, a)$ is convex, and stays as close as possible below $g(\mathbf{x}, a)$. Such a matrix $\mathbf{D}$ can be found by minimizing the maximum distance between $g_L(\mathbf{x}, a)$ and $g(\mathbf{x}, a)$, and comes out as the solution of the following minimization problem:

$$\min_{\mathbf{D}} \ (\mathbf{x}_U - \mathbf{x}_L)^T \mathbf{D}(\mathbf{x}_U - \mathbf{x}_L) \ \text{ s. to } \ \mathbf{H} + 2\mathbf{D} \succeq \mathbf{0} \tag{15}$$

where the constraint on the Hessian ensures that $g_L(\mathbf{x}, a)$ remains convex. Since (15) is a semi-definite program, it can be solved efficiently using available convex optimization software; e.g., the interior point optimization routine in SeDuMi [27]. Having selected $\mathbf{D}$ as in (15), $\min_{\mathbf{x} \in \mathcal{X}_1(\mu), \mathbf{x}_L \le \mathbf{x} \le \mathbf{x}_U} g_L(\mathbf{x}, a)$ is a convex problem (quadratic cost under linear constraints); thus, similar to the upper bound $\mathcal{U}$, the lower bound $\mathcal{L}$ on $g_{\text{box}}^\star(a)$ can be obtained efficiently.

The detailed inner loop (BB scheme) is tabulated as Algorithm 1-a. It amounts to successively splitting the initial box $-\mu\mathbf{1} \leq \mathbf{x} \leq \mu\mathbf{1}$, which is the smallest one containing $\mathcal{X}_1(\mu)$. Per inner iteration $i$, variable $\mathcal{U}$ keeps track of the upper bound on $g_{\text{box}}^\star(a)$, which at the end outputs to the outer loop the nearest estimate of $g^\star(a)$. Concurrently, the lower bound $\mathcal{L}$ on $g_{\text{box}}^\star(a)$ determines whether the current box needs to be further split, or discarded, if the difference $\mathcal{U} - \mathcal{L}$ is smaller than the pre-selected margin $\delta$. This iterative splitting leads to a decreasing $\mathcal{U}$ and a tighter $\mathcal{L}$, both of which prevent further splitting.

Recapitulating, the outer bisection-based iteration tabulated as Algorithm 1-b calls Algorithm 1-a to find a feasible $\delta$-optimal solution $\mathbf{x}_g^\star$ to evaluate the sign of $g^\star(a)$ in (12). Since $\mathbf{x}_g^\star$ is not the exact global minimum of (12), positivity of $g(\mathbf{x}_g^\star, a)$ does not necessarily imply $g^\star(a) > 0$. But $\mathbf{x}_g^\star$ is $\delta$-optimal, meaning that $g^\star(a) \geq g(\mathbf{x}_g^\star, a) - \delta$; thus, $g(\mathbf{x}_g^\star, a) > \delta$, in which case the lower bound $l_{i+1}$ is updated to $(l_i + u_i)/2$; otherwise, if $g(\mathbf{x}_g^\star, a) \in (0, \delta)$, then $l_{i+1}$ should be set to $(l_i + u_i)/2 - \delta$.

As far as convergence is concerned, the following result can be established.

**Proposition 2:** *($\varepsilon$-optimal convergence) After at most $\left\lceil \ln\left(\frac{u_0}{\varepsilon - 2\delta}\right) / \ln(2) \right\rceil$ iterations, Algorithm 1-b outputs an $\varepsilon$-optimal solution $\mathbf{x}_\varepsilon^\star$ to* (10); *that is,*

$$\mathbf{x}_\varepsilon^\star \in \mathcal{X}_1(\mu), \quad \text{and} \quad a^\star \leq f(\mathbf{x}_\varepsilon^\star) \leq a^\star + \varepsilon. \tag{16}$$

*Proof:* Upon updating the lower and upper bounds, it holds per outer iteration $i \geq 1$ that $u_i - l_i \leq \frac{1}{2}(u_{i-1} - l_{i-1}) + \delta$; and by induction, $u_i - l_i \leq \left(\frac{1}{2}\right)^i u_0 + 2\delta$, when $l_0 = 0$. The latter implies that if the number of iterations $i \geq \left\lceil \ln\left(\frac{u_0}{\varepsilon - 2\delta}\right) / \ln(2) \right\rceil$, the distance $u_i - l_i \leq \varepsilon$ is satisfied.

Since per outer iteration Algorithm 1-a outputs $\mathbf{x}_g^\star \in \mathcal{X}_1(\mu)$, it holds that the updated $\mathbf{x}_\varepsilon^\star$ is also feasible. Further, the bisection process guarantees that $l_i \leq a^\star \leq f(\mathbf{x}_\varepsilon^\star) \leq u_i$ per iteration $i$. Since Algorithm 1-b ends with $u_i - l_i \leq \varepsilon$, the inequality in (16) follows readily. ∎

Proposition 2 quantifies the number of outer iterations needed by the bisection-based Algorithm 1-b to approach within $\varepsilon$ the global optimum of (10). In addition, the inner (BB) iterations bounding $g_{\text{box}}^\star(a)$ are expected to be fast converging because the box function in (13) is tailored for the box constraints induced by the $\ell_1$-norm regularization. Nonetheless, similar to all BB algorithms, the complexity of Algorithm 1-a does not have guaranteed polynomial complexity on average. The latter necessitates as few calls of Algorithm 1-a, which means as few outer iterations. Proposition 2 reveals that critical to this end is the initial upper bound $u_0$ (Algorithm 1-b simply initializes with $u_0 = f(\mathbf{0})$).

This motivates the efficient suboptimal S-TLS solver of the next subsection, which is of paramount importance not only on its own, but also for initializing the $\varepsilon$-optimal algorithm.

*B. Alternating Descent Sub-Optimal Algorithm*

The starting point for a computationally efficient S-TLS solver is the formulation in (5). Given $\mathbf{E}$, the cost in (5) has the form of the Lasso problem in (1); while given $\mathbf{x}$, it reduces to a quadratic form, which admits closed-form solution wrt $\mathbf{E}$. These observations suggest an iterative block coordinate descent algorithm yielding successive estimates of $\mathbf{x}$ with $\mathbf{E}$ fixed, and alternately of $\mathbf{E}$ with $\mathbf{x}$ fixed. Specifically, with the iterate $\mathbf{E}(i)$ given per iteration $i \geq 0$, the iterate $\mathbf{x}(i)$ is obtained by solving the Lasso-like convex problem as [cf. (1)]

$$\mathbf{x}(i) = \arg\min_{\mathbf{x}} \|\mathbf{y} - [\mathbf{A} + \mathbf{E}(i)]\mathbf{x}\|_2^2 + \lambda\|\mathbf{x}\|_1 . \tag{17}$$

With $\mathbf{x}(i)$ available, $\mathbf{E}(i+1)$ for the ensuing iteration is found as

$$\mathbf{E}(i+1) = \arg\min_{\mathbf{E}} \|\mathbf{y} - \mathbf{A}\mathbf{x}(i) - \mathbf{E}\mathbf{x}(i)\|_2^2 + \|\mathbf{E}\|_F^2. \tag{18}$$

By setting the first-order derivative of the cost wrt $\mathbf{E}$ equal to zero, the optimal solution to the quadratic problem (18) is obtained in closed form as

$$\mathbf{E}(i+1) = (1 + \|\mathbf{x}(i)\|_2^2)^{-1}[\mathbf{y} - \mathbf{A}\mathbf{x}(i)]\mathbf{x}^T(i). \tag{19}$$

The iterations are initialized at $i = 0$ by setting $\mathbf{E}(0) = \mathbf{0}_{m \times n}$. Substituting the latter into (17), yields $\mathbf{x}(0) = \hat{\mathbf{x}}_{Lasso}$ in (1). That this is a good initial estimate is corroborated by the result in [20], which shows that even with perturbations present in both $\mathbf{A}$ and $\mathbf{y}$, the CS (and thus Lasso) estimators yield accurate reconstruction. In view of the fact that the block coordinate descent iterations ensure that the cost in (5) is non-increasing, the final estimates upon convergence will be at least as accurate.

The block coordinate descent algorithm is provably convergent to a stationary point of the S-TLS cost in (5), and thus to its equivalent forms in (3), (6) and (10), as asserted in the following proposition.

**Proposition 3:** *(Convergence of alternating descent) Given arbitrary initialization, the iterates $\{\mathbf{E}(i), \mathbf{x}(i)\}$ given by* (17) *and* (19) *converge monotonically at least to a stationary point of the S-TLS problem* (3).

*Proof:* The argument relies on the basic convergence result in [29]. The alternating descent algorithm specified by (17) and (19) is a special case of the block coordinate descent method using the cyclic rule for minimizing the cost in (5). The first two summands of this cost are differentiable wrt the optimization variables, while the non-differential third term ($\ell_1$-norm regularization) is separable in the entries of $\mathbf{x}$. Hence, the three summands satisfy the assumptions (B1)–(B3) and (C2) in [29]. Convergence of the iterates $\{\mathbf{E}(i), \mathbf{x}(i)\}$ to a coordinate minimum point of the cost thus follows by appealing to [29, Thm. 5.1]. Moreover, the first summand is Gâteaux-differentiable over its domain which is open. Hence, the cost in (5) is regular at each coordinate's minimum point, and every coordinate's minimum point becomes a stationary point; see [29, Lemma 3.1]. Monotonicity of the convergence follows simply because the cost per iteration may either reduce or maintain its value. ∎

Proposition 3 solidifies the merits of the alternating descent S-TLS solver. Simulated tests will further demonstrate that the local optimum guaranteed by this computationally efficient scheme is very close to the global optimum attained by the more complex scheme of the previous subsection.

Since estimating $\mathbf{E}$ is simple using the closed form in (18), it is useful at this point to explore modifications, extensions and tailored solvers for the problem in (17) by adapting to the present setup existing results from the Lasso literature dealing with problem (1). From the plethora of available options to solve (17), it is worth mentioning two computationally efficient ones: the least-angle regression (LARS), and the coordinate descent (CD); see e.g., [19]. LARS provides the entire "solution path" of (17) for all $\lambda > 0$ at complexity comparable to LS. On the other hand, if a single "best" value of $\lambda$ is fixed using the cross-validation scheme [25], then CD is the state-of-the-art choice for solving (17).

CD in the present context cycles between iterates $\mathbf{E}(i)$, and scalar iterates of the $\mathbf{x}(i)$ entries. Suppose that the $\nu$-th entry $x_\nu(i)$ is to be found. Precursor entries $\{x_1(i), \ldots, x_{\nu-1}(i)\}$ have been already obtained in the $i$-th iteration, and postcursor entries $\{x_{\nu+1}(i-1), \ldots, x_n(i-1)\}$ are also available from the previous $(i-1)$-st iteration along with $\mathbf{E}(i)$ obtained in closed form as in (19). If $\boldsymbol{\alpha}_\nu(i)$ denotes the $\nu$-th column of $[\mathbf{A} + \mathbf{E}(i)]$, the effect of these known entries can be removed from $\mathbf{y}$ by forming

$$\mathbf{e}_\nu(i) := \mathbf{y} - \sum_{j=1}^{\nu-1} \boldsymbol{\alpha}_j(i) x_j(i) - \sum_{j=\nu+1}^{n} \boldsymbol{\alpha}_j(i) x_j(i-1) \ . \qquad (20)$$

Using (20), the vector optimization problem in (17) reduces to the following scalar one with $x_\nu(i)$ as unknown: $x_\nu(i) = \arg\min_{x_\nu}[\|\mathbf{e}_\nu(i) - \boldsymbol{\alpha}_\nu(i) x_\nu\|_2^2 + \lambda |x_\nu|]$. This *scalar* Lasso problem is known to admit a closed-form solution expressed in terms of a soft thresholding operator (see e.g., [19])

$$x_\nu(i) = \operatorname{sign}\left(\mathbf{e}_\nu^T(i) \boldsymbol{\alpha}_\nu(i)\right) \left[\frac{\mathbf{e}_\nu^T(i) \boldsymbol{\alpha}_\nu(i)}{\|\boldsymbol{\alpha}_\nu(i)\|_2^2} - \frac{\lambda}{2\|\boldsymbol{\alpha}_\nu(i)\|_2^2}\right]_+, \quad \nu = 1, \ldots, n \qquad (21)$$

where $\operatorname{sign}(\cdot)$ denotes the sign operator, and $[\chi]_+ := \chi$, if $\chi > 0$, and zero otherwise.

Cycling through the closed forms (19)-(21) explains why CD here is faster than, and thus preferable over general-purpose convex optimization solvers of (17). Another factor contributing to its speed is the sparsity of $\mathbf{x}(i)$, which implies that starting up with the all-zero vector, namely $\mathbf{x}(-1) = \mathbf{0}_{n \times 1}$, offers initialization close to a stationary point of the cost in (5). Convergence to this stationary point is guaranteed by using the results in [29], along the lines of Proposition 3. Note also that larger values of $\lambda$ in (21) force more entries of $\mathbf{x}(i)$ to be shrunk to zero, which corroborates the role of $\lambda$ as a sparsity-tuning parameter. The CD based S-TLS solver is tabulated as Algorithm 2.

**Remark 1:** *(Regularization options for S-TLS)* Lasso estimators are known to be biased, but modifications are available to remedy bias effects. One such modification is the weighted Lasso, which replaces the $\ell_1$-norm in (3) by its weighted version, namely $\sum_{\nu=1}^{n} w_\nu |x_\nu|$, where the weights $\{w_\nu\}$ are chosen using the LS solution [31]. An alternative popular choice is to replace the $\ell_1$-norm with concave regularization terms [15], such as $\sum_{\nu=1}^{n} \log(x_\nu + \delta_1)$, where $\delta_1$ is a small positive constant introduced to avoid numerical instability. In addition to mitigating bias effects, concave regularization terms provide tighter approximations to the $\ell_0$-(pseudo)norm, and although they render the cost in (3) nonconvex, they are known to converge very fast to an improved estimate of $\mathbf{x}$, when initialized with the Lasso solution [15].

**Remark 2:** *(Group Lasso and Matrix S-TLS)* When groups $\{\mathbf{x}_g\}_{g=1}^{G}$ of $\mathbf{x}$ entries are a priori known to be zero or nonzero (as a group), the $\ell_1$-norm in (3) must be replaced by the sum of $\ell_2$-norms, namely $\sum_{g=1}^{G} \|\mathbf{x}_g\|_2$. The resulting group S-TLS estimate can be obtained using the group-Lasso solver [19]. In the present context, this is further useful if one considers the matrix counterpart of the S-TLS problem in (3), which in its unconstrained form can be written as [cf. (5)]

$$\left\{\hat{\mathbf{X}}_{S-TLS}, \hat{\mathbf{E}}_{S-TLS}\right\} = \arg\min_{\mathbf{X},\mathbf{E}} \left[\|\mathbf{Y} - (\mathbf{A} + \mathbf{E})\mathbf{X}\|_F^2 + \|\mathbf{E}\|_F^2 + \lambda \sum_{\nu=1}^{n} \|\mathbf{x}_\nu^T\|_2\right] \quad (22)$$

where $\mathbf{x}_\nu^T$ denotes the $\nu$-th row of the $n \times L$ unknown matrix $\mathbf{X}$, which is sparse in the sense that a number of its rows are zero, and has to be estimated using an $m \times L$ data matrix $\mathbf{Y}$ along with the regression matrix $\mathbf{A}$, both with perturbations present. Problem (22) can be solved using block coordinate descent cycling between iterates $\mathbf{E}(i)$ and rows $\mathbf{x}_\nu^T(i)$ as opposed to scalar entries as in (21).

## V. WEIGHTED AND STRUCTURED S-TLS

Apart from the optimality links established in Proposition 1 under (as1), the S-TLS criteria in (3), (5), and (6) make no assumption on the perturbations $[\mathbf{E}\ \mathbf{e}]$. In this sense, the S-TLS solvers of the previous section find universal applicability. However, one expects that exploiting prior information on $[\mathbf{E}\ \mathbf{e}]$, can only lead to improved performance. Thinking for instance along the lines of weighted LS, one is motivated to *weight* $\|\mathbf{E}\|_F^2$ and $\|\mathbf{e}\|_2^2$ in (5) by the inverse covariance matrix of $\mathbf{E}$ and $\mathbf{e}$, respectively, whenever those are known and are not both equal to $\mathbf{I}$. As a second motivating example, normal equations, involved in e.g., linear prediction, entail *structure* in $\mathbf{E}$ and $\mathbf{e}$ that capture sample estimation errors present in the matrix $[\mathbf{A}\ \mathbf{y}]$, which is Toeplitz. Prompted by these examples, this section is about broadening the scope of S-TLS with weighted and structured forms capitalizing on prior information available about the matrix $[\mathbf{E}\ \mathbf{e}]$. To this end, it is prudent to quantify first the notion of structure.

**Definition 1.** *The $m \times (n+1)$ data matrix $[\mathbf{A}\ \mathbf{y}](\mathbf{p})$ has structure characterized by an $n_p \times 1$ parameter vector $\mathbf{p}$, if and only if there is a mapping such that $\mathbf{p} \in \mathbb{R}^{n_p} \to [\mathbf{A}\ \mathbf{y}](\mathbf{p}) := \mathbf{S}(\mathbf{p}) \in \mathbb{R}^{m \times (n+1)}$.*

Definition 1 is general enough to encompass any (even unstructured) matrix $[\mathbf{A}\ \mathbf{y}](\mathbf{p})$, by simply letting $\mathbf{p} := \text{vec}([\mathbf{A}\ \mathbf{y}]) \in \mathbb{R}^{m(n+1)}$ comprise all entries of $[\mathbf{A}\ \mathbf{y}]$. However, it becomes more relevant when $n_p \ll m(n+1)$, the case in which $\mathbf{p}$ characterizes $[\mathbf{A}\ \mathbf{y}]$ parsimoniously. Application examples are abundant: structure in Toeplitz and Hankel matrices encountered with system identification, deconvolution, and linear prediction; as well as in circulant and Vandermonde matrices showing up in spatio-temporal harmonic retrieval problems [23]. Structured matrices $\mathbf{A}$ and sparse vectors $\mathbf{x}_o$ emerge also in contemporary CS gridding-based applications e.g., for spectral analysis and estimation of time-varying channels, where rows of the FFT matrix are selected at random. (This last setting appears when training orthogonal frequency-division multiplexing (OFDM) input symbols are used to estimate communication links exhibiting variations due to mobility-induced Doppler effects [6].)

Consider now re-casting the S-TLS criteria in terms of $\mathbf{p}$, and its associated perturbation vector denoted by $\boldsymbol{\epsilon} \in \mathbb{R}^{n_p}$. The Frobenius norm in the cost of (3a) is mapped to the $\ell_2$-norm of $\boldsymbol{\epsilon}$; and to allow for weighting the structured perturbation vector using a symmetric positive definite matrix $\mathbf{W} \in \mathbb{R}^{n_p \times n_p}$, the weighted counterpart of $\|[\mathbf{E}\ \mathbf{e}]\|_F^2$ becomes $\boldsymbol{\epsilon}^T \mathbf{W} \boldsymbol{\epsilon}$. With regards to the constraint, recall first from Definition 1 that $\mathbf{S}(\mathbf{p}) = [\mathbf{A}\ \mathbf{y}]$, which implies $\mathbf{S}(\mathbf{p}+\boldsymbol{\epsilon}) = [\mathbf{A}+\mathbf{E}\ \mathbf{y}+\mathbf{e}]$; hence, re-writing (3b) as $[\mathbf{A}+\mathbf{E}\ \mathbf{y}+\mathbf{e}]\left[\mathbf{x}^T, -1\right]^T = \mathbf{0}$, yields the structured constraint as $\mathbf{S}(\mathbf{p}+\boldsymbol{\epsilon})\left[\mathbf{x}^T, -1\right]^T = \mathbf{0}$. Putting things together, leads to the combined weighted-structured S-TLS version of (3) as

$$\min_{\mathbf{x}, \boldsymbol{\epsilon}}\ \boldsymbol{\epsilon}^T \mathbf{W} \boldsymbol{\epsilon} + \lambda \|\mathbf{x}\|_1 \tag{23a}$$

$$\text{s. to}\ \mathbf{S}(\mathbf{p}+\boldsymbol{\epsilon}) \begin{bmatrix} \mathbf{x} \\ -1 \end{bmatrix} = \mathbf{0} \tag{23b}$$

which clearly subsumes the structure-only form as a special case corresponding to $\mathbf{W} = \mathbf{I}$.

To confine the structure quantified in Definition 1, two conditions will be imposed, which are commonly adopted by TLS approaches [23], and are satisfied by most applications mentioned so far.

**(as2)** *The structure mapping in Definition 1 is separable, meaning that with $\mathbf{p} = [(\mathbf{p}^A)^T\ (\mathbf{p}^y)^T]^T$, where $\mathbf{p}^A \in \mathbb{R}^{n_A}$ and $\mathbf{p}^y \in \mathbb{R}^{n_y}$, it holds that $\mathbf{S}(\mathbf{p}) := [\mathbf{A}\ \mathbf{y}](\mathbf{p}) = [\mathbf{A}(\mathbf{p}^A)\ \mathbf{y}(\mathbf{p}^y)]$. In addition, the separable structure mapping is linear (more precisely affine), if and only if the $\mathbf{S}(\mathbf{p})$ matrix is composed of known structural elements, namely "matrix atoms" $\mathbf{S}_0$, $\{\mathbf{S}_k^A\}_{k=1}^{n_A}$ and "vector atoms" $\{\mathbf{s}_k^y\}_{k=1}^{n_y}$, so that*

$$\mathbf{S}(\mathbf{p}) = \mathbf{S}_0 + \left[\sum_{k=1}^{n_A} p_k^A \mathbf{S}_k^A \quad \sum_{k=1}^{n_y} p_k^y \mathbf{s}_k^y\right] \tag{24}$$

where $p_k^A$ ($p_k^y$) denotes the k-th entry of $\mathbf{p}^A$ ($\mathbf{p}^y$).

Similar to Definition 1, (24) is general enough to encompass even unstructured matrices $\mathbf{S}(\mathbf{p}) := [\mathbf{A}\,\mathbf{y}]$, by setting $\mathbf{S}_0 = \mathbf{0}$, $\mathbf{p} := \text{vec}([\mathbf{A}\,\mathbf{y}]) \in \mathbb{R}^{m(n+1)}$, $n_p := n_A + n_y = mn + m$, and selecting the $m$ vector atoms ($mn$ matrix atoms) as the canonical vectors (matrices), each with one entry equal to 1 and all others equal to 0. Again, interesting structures are those with $n_A \ll mn$ and/or $n_y \ll m$. (Consider for instance a circulant $m \times n$ matrix $\mathbf{A}$, which can be represented as in (24) using $n_A = m$ matrix atoms.)

Separability and linearity will turn out to simplify the constraint in (23b) for some given matrix atoms and vector atoms collected for notational brevity in the matrices

$$\boldsymbol{S}^A := [\mathbf{S}_1^A \cdots \mathbf{S}_{n_A}^A] \quad \text{and} \quad \boldsymbol{S}^y := [\mathbf{s}_1^y \cdots \mathbf{s}_{n_y}^y] \in \mathbb{R}^{m \times n_y} . \tag{25}$$

Indeed, linearity in (as2) allows one to write $\mathbf{S}(\mathbf{p} + \boldsymbol{\epsilon}) = \mathbf{S}(\mathbf{p}) + \mathbf{S}(\boldsymbol{\epsilon})$, and the constraint (23b) as: $\mathbf{S}(\boldsymbol{\epsilon})[\mathbf{x}^T, -1]^T = -\mathbf{S}(\mathbf{p})[\mathbf{x}^T, -1]^T = \mathbf{y} - \mathbf{A}\mathbf{x}$; while separability implies that $\mathbf{S}(\boldsymbol{\epsilon})[\mathbf{x}^T, -1]^T = [\sum_{k=1}^{n_A} \epsilon_k^A \mathbf{S}_k^A \quad \sum_{k=1}^{n_y} \epsilon_k^y \mathbf{s}_k^y][\mathbf{x}^T, -1]^T = \boldsymbol{S}^A(\mathbf{I} \otimes \mathbf{x})\boldsymbol{\epsilon}^A - \boldsymbol{S}^y \boldsymbol{\epsilon}^y$, where the definitions $\boldsymbol{\epsilon} := [(\boldsymbol{\epsilon}^A)^T \, (\boldsymbol{\epsilon}^y)^T]^T$ and (25) were used in the last equality along with the identity $\sum_{k=1}^{n_A} \epsilon_k^A \mathbf{S}_k^A \mathbf{x} = \boldsymbol{S}^A (\mathbf{I} \otimes \mathbf{x}) \boldsymbol{\epsilon}^A$. In a nutshell, (23b) under (as2) becomes $\boldsymbol{S}^A(\mathbf{I} \otimes \mathbf{x})\boldsymbol{\epsilon}^A - \boldsymbol{S}^y \boldsymbol{\epsilon}^y = \mathbf{y} - \mathbf{A}\mathbf{x}$, in which $\boldsymbol{\epsilon}^A$ is decoupled from $\boldsymbol{\epsilon}^y$.

Therefore, the weighted and structured (WS)S-TLS problem in (23) reduces to [cf. (3)]

$$\min_{\mathbf{x}, \boldsymbol{\epsilon}^A, \boldsymbol{\epsilon}^y} \begin{bmatrix} \boldsymbol{\epsilon}^A \\ \boldsymbol{\epsilon}^y \end{bmatrix}^T \mathbf{W} \begin{bmatrix} \boldsymbol{\epsilon}^A \\ \boldsymbol{\epsilon}^y \end{bmatrix} + \lambda \|\mathbf{x}\|_1 \tag{26a}$$

$$\text{s. to} \quad [\boldsymbol{S}^A(\mathbf{I} \otimes \mathbf{x}) \; -\boldsymbol{S}^y] \begin{bmatrix} \boldsymbol{\epsilon}^A \\ \boldsymbol{\epsilon}^y \end{bmatrix} = \mathbf{y} - \mathbf{A}\mathbf{x} \tag{26b}$$

or in a more compact form as: $\min_{\mathbf{x}, \boldsymbol{\epsilon}} \{\boldsymbol{\epsilon}^T \mathbf{W} \boldsymbol{\epsilon} + \lambda \|\mathbf{x}\|_1\}$ s.to $\boldsymbol{G}(\mathbf{x})\boldsymbol{\epsilon} = \mathbf{r}(\mathbf{x})$, after defining

$$\boldsymbol{G}(\mathbf{x}) := [\boldsymbol{S}^A(\mathbf{I} \otimes \mathbf{x}) \; \boldsymbol{S}^y] \quad \text{and} \quad \mathbf{r}(\mathbf{x}) := \mathbf{y} - \mathbf{A}\mathbf{x} . \tag{27}$$

Comparing (3) with (26) allows one to draw apparent analogies: both involve three sets of optimization variables, and both are nonconvex because two of these sets enter the corresponding constraints in a bilinear fashion [cf. product of $\mathbf{E}$ with $\mathbf{x}$ in (3b), and $\boldsymbol{\epsilon}^A$ with $\mathbf{x}$ in (26b)].

Building on these analogies, the following lemma shows how to formulate WSS-TLS criteria, paralleling those of Lemma 1, where one or two sets of variables were eliminated to obtain efficient, provably convergent solvers, and establish statistical optimality links within the EIV model in (4).

**Lemma 2:** *The constrained WSS-TLS form in* (23) *is equivalent to two unconstrained nonconvex*

*optimization problems: (a) one involving* $\mathbf{x}$ *and* $\boldsymbol{\epsilon}^A$ *variables, namely*

$$\{\hat{\mathbf{x}}, \hat{\boldsymbol{\epsilon}}^A\}_{WSS-TLS} = \arg\min_{\mathbf{x}, \boldsymbol{\epsilon}^A} \begin{bmatrix} \boldsymbol{\epsilon}^A \\ (\boldsymbol{S}^y)^\dagger [\boldsymbol{S}^A(\mathbf{I} \otimes \mathbf{x})\boldsymbol{\epsilon}^A - \mathbf{r}(\mathbf{x})] \end{bmatrix}^T \mathbf{W} \begin{bmatrix} \boldsymbol{\epsilon}^A \\ (\boldsymbol{S}^y)^\dagger [\boldsymbol{S}^A(\mathbf{I} \otimes \mathbf{x})\boldsymbol{\epsilon}^A - \mathbf{r}(\mathbf{x})] \end{bmatrix}$$
$$+ \lambda \|\mathbf{x}\|_1 \quad (28)$$

*where* $\boldsymbol{S}^y$ *is assumed full rank and square*[1], *i.e.,* $m = n_y$ *in* (27); *and also (b) one involving only the variable* $\mathbf{x}$, *expressed using the definitions in* (27), *as*

$$\hat{\mathbf{x}}_{WSS-TLS} = \arg\min_{\mathbf{x}} \{\mathbf{r}^T(\mathbf{x}) \left[\boldsymbol{G}(\mathbf{x}) \mathbf{W}^{-1} \boldsymbol{G}^T(\mathbf{x})\right]^\dagger \mathbf{r}(\mathbf{x}) + \lambda \|\mathbf{x}\|_1\}. \quad (29)$$

*Proof:* Constraint (26b) can be solved uniquely for $\boldsymbol{\epsilon}^y$ to obtain $\boldsymbol{\epsilon}^y = (\boldsymbol{S}^y)^\dagger[\boldsymbol{S}^A(\mathbf{I}\otimes\mathbf{x})\boldsymbol{\epsilon}^A - (\mathbf{y} - \mathbf{A}\mathbf{x})]$. Plug the latter with the definition of $\mathbf{r}(\mathbf{x})$ from (27) into the quadratic form in (26a) to recognize that (26) is equivalent to the unconstrained form in (28) with the $\boldsymbol{\epsilon}^y$ variable eliminated.

To arrive at (29), suppose that $\mathbf{x}$ is given and view the compact form of (26) (after ignoring $\lambda\|\mathbf{x}\|_1$) as the following weighted minimum-norm LS problem: $\min_{\mathbf{x}, \boldsymbol{\epsilon}}\{\boldsymbol{\epsilon}^T\mathbf{W}\boldsymbol{\epsilon} + \lambda\|\mathbf{x}\|_1\}$ s.to $\boldsymbol{G}(\mathbf{x})\boldsymbol{\epsilon} = \mathbf{r}(\mathbf{x})$. Solving the latter in closed form expresses $\boldsymbol{\epsilon}$ in terms of $\mathbf{x}$ as: $\boldsymbol{\epsilon}(\mathbf{x}) = \mathbf{W}^{-1}\boldsymbol{G}^T(\mathbf{x})\left[\boldsymbol{G}(\mathbf{x})\mathbf{W}^{-1}\boldsymbol{G}^T(\mathbf{x})\right]^\dagger \mathbf{r}(\mathbf{x})$. Substitute now $\boldsymbol{\epsilon}(\mathbf{x})$ back into the cost, and reinstate $\lambda\|\mathbf{x}\|_1$, to obtain (29). ∎

The formulation in (28) suggests directly an iterative WSS-TLS solver based on the block coordinate descent method. Specifically, suppose that the estimate $\boldsymbol{\epsilon}^A(i)$ of $\boldsymbol{\epsilon}^A$ is available at iteration $i$. Substituting $\boldsymbol{\epsilon}^A(i)$ into (28), allows estimating $\mathbf{x}$ as

$$\mathbf{x}(i) = \arg\min_{\mathbf{x}} \begin{bmatrix} \boldsymbol{\epsilon}^A(i) \\ (\boldsymbol{S}^y)^\dagger[\boldsymbol{S}^A(\mathbf{I}\otimes\mathbf{x})\boldsymbol{\epsilon}^A(i) - \mathbf{r}(\mathbf{x})] \end{bmatrix}^T \mathbf{W} \begin{bmatrix} \boldsymbol{\epsilon}^A(i) \\ (\boldsymbol{S}^y)^\dagger[\boldsymbol{S}^A(\mathbf{I}\otimes\mathbf{x})\boldsymbol{\epsilon}^A(i) - \mathbf{r}(\mathbf{x})] \end{bmatrix} + \lambda\|\mathbf{x}\|_1. \quad (30)$$

Since $\mathbf{r}(\mathbf{x})$ is linear in $\mathbf{x}$ [cf. (27)], the cost in (30) is convex (quadratic regularized by the $\ell_1$-norm as in the Lasso cost in (1)); thus, it can be minimized efficiently. Likewise, given $\mathbf{x}(i)$ the perturbation vector for the ensuing iteration can be found in closed form since the pertinent cost is quadratic; that is,

$$\boldsymbol{\epsilon}^A(i+1) = \arg\min_{\boldsymbol{\epsilon}^A} \begin{bmatrix} \boldsymbol{\epsilon}^A \\ (\boldsymbol{S}^y)^\dagger[\boldsymbol{S}^A(\mathbf{I}\otimes\mathbf{x}(i))\boldsymbol{\epsilon}^A - \mathbf{r}(\mathbf{x}(i))] \end{bmatrix}^T \mathbf{W} \begin{bmatrix} \boldsymbol{\epsilon}^A \\ (\boldsymbol{S}^y)^\dagger[\boldsymbol{S}^A(\mathbf{I}\otimes\mathbf{x}(i))\boldsymbol{\epsilon}^A - \mathbf{r}(\mathbf{x}(i))] \end{bmatrix}.$$
$$(31)$$

---

[1] Tall $\boldsymbol{S}^y$ matrices with full column rank can be handled too for block diagonal weight matrices $\mathbf{W}$ typically adopted with separable structures. This explains why the pseudo-inverse of $\boldsymbol{S}^y$ is used in this section instead of its inverse; but exposition of the proof simplifies considerably for the square case. Note also that the full rank assumption is not practically restrictive because data matrices perturbed by noise of absolutely continuous pdf have full rank almost surely.

To express $\boldsymbol{\epsilon}^A(i+1)$ compactly, partition $\mathbf{W}$ in accordance with $\mathbf{p} = [(\mathbf{p}^A)^T, \ (\mathbf{p}^y)^T]^T$; i.e., let

$$\mathbf{W} = \begin{bmatrix} \mathbf{W}_{AA} & \mathbf{W}_{Ay} \\ \mathbf{W}_{Ay}^T & \mathbf{W}_{yy} \end{bmatrix}. \tag{32}$$

Using (32), and equating to zero the gradient (wrt $\boldsymbol{\epsilon}^A$) of the cost in (31), yields the closed form

$$\boldsymbol{\epsilon}^A(i+1) = \left\{ \check{\mathbf{S}}(\mathbf{x}(i))\mathbf{W}\check{\mathbf{S}}^T(\mathbf{x}(i)) \right\}^\dagger \check{\mathbf{S}}(\mathbf{x}(i))[\mathbf{W}_{Ay}^T, \ \mathbf{W}_{yy}^T]^T (\boldsymbol{S}^y)^\dagger \mathbf{r}(\mathbf{x}(i)) \tag{33}$$

where $\check{\boldsymbol{S}}(\mathbf{x}(i)) := \left[ \mathbf{I}, \ [(\boldsymbol{S}^y)^\dagger \boldsymbol{S}^A(\mathbf{I} \otimes \mathbf{x}(i))]^T \right]$.

Initialized with $\boldsymbol{\epsilon}^A(0) = \mathbf{0}_{n_A \times 1}$, the algorithm cycles between iterations (30) and (33). Mimicking the steps of Proposition 3, it is easy to show that these iterations are convergent as asserted in the following.

**Proposition 4:** *(Convergence). The iterates in* (30) *and* (33) *converge monotonically at least to a stationary point of the cost in* (23)*, provided that* $\boldsymbol{S}^y$ *in* (27) *has full column rank.*

As with the solver of Section IV-B, CD is also applicable to the WSS-TLS solver, by cycling between $\boldsymbol{\epsilon}^A(i)$ and scalar iterates of the $\mathbf{x}(i)$ entries. To update the $\nu$-th entry $x_\nu(i)$, suppose precursor entries $\{x_1(i), \ldots, x_{\nu-1}(i)\}$ have been already obtained in the $i$-th iteration, and postcursor entries $\{x_{\nu+1}(i-1), \ldots, x_n(i-1)\}$ are also available from the previous $(i-1)$-st iteration along with $\boldsymbol{\epsilon}^A(i)$, found in closed form as in (33). Letting $\boldsymbol{\alpha}_\nu(i)$ denote the $\nu$-th column of $[(\boldsymbol{S}^y)^\dagger (\mathbf{A} + \sum_{k=1}^{n_A} \epsilon_k^A \mathbf{S}_k^A)]$, the effect of these known entries can be removed from $\mathbf{y}$ by forming [cf. (20)]

$$\mathbf{e}_\nu(i) := (\boldsymbol{S}^y)^\dagger \mathbf{y} - \sum_{j=1}^{\nu-1} \boldsymbol{\alpha}_j(i) x_j(i) - \sum_{j=\nu+1}^n \boldsymbol{\alpha}_j(i) x_j(i-1) . \tag{34}$$

Using (34), the vector optimization problem in (30) now reduces to the following scalar one with $x_\nu(i)$ as unknown: $x_\nu(i) = \arg\min_{x_\nu} \{ \|\boldsymbol{\alpha}_\nu(i) x_\nu - \mathbf{e}_\nu(i)\|_{\mathbf{W}_{yy}}^2 + 2[\boldsymbol{\epsilon}^A(i)]^T \mathbf{W}_{Ay} \boldsymbol{\alpha}_\nu(i) x_\nu + \lambda |x_\nu| \}$, where $\|\cdot\|_{\mathbf{W}_{yy}}$ denotes the $\ell_2$-norm weighted by $\mathbf{W}_{yy}$. The solution of this scalar Lasso problem can be expressed using the same soft-thresholding form as in (21), and is given by

$$x_\nu(i) = \text{sign}\left([\mathbf{e}_\nu(i)\mathbf{W}_{yy} - \boldsymbol{\epsilon}^A(i)\mathbf{W}_{Ay}^T]^T \boldsymbol{\alpha}_\nu(i)\right) \left[ \frac{[\mathbf{e}_\nu(i)\mathbf{W}_{yy} - \boldsymbol{\epsilon}^A(i)\mathbf{W}_{Ay}^T]^T \boldsymbol{\alpha}_\nu(i)}{\|\boldsymbol{\alpha}_\nu(i)\|_{\mathbf{W}_{yy}}^2} - \frac{\lambda}{2\|\boldsymbol{\alpha}_\nu(i)\|_{\mathbf{W}_{yy}}^2} \right]_+ . \tag{35}$$

This block CD algorithm enjoys fast convergence (at least) to a stationary point, thanks both to the simplicity of (35), and the sparsity of $\mathbf{x}(i)$.

The WSS-TLS criterion in (28) is also useful to establish its statistical optimality under a structured EIV model, with output-input data obeying the relationships

$$\mathbf{y} = \mathbf{A}(\mathbf{p}_o^A)\mathbf{x}_o + (-\boldsymbol{S}^y \boldsymbol{\epsilon}_y), \quad \mathbf{A} = \mathbf{A}(\mathbf{p}_o^A) + \left( -\sum_{k=1}^{n_A} \epsilon_{A,k} \mathbf{S}_k^A \right) \tag{36}$$

where perturbation vectors $\epsilon_A$ and $\epsilon_y$ play the role of $\mathbf{E}_A$ and $\mathbf{e}_y$ in (4), and differ from the optimization variables $\epsilon^A$ and $\epsilon^y$ in (26). Unknown are the vector $\mathbf{x}_o$, and the inaccessible input matrix $\mathbf{A}(\mathbf{p}_o^A)$, characterized by the vector $\mathbf{p}_o^A$. The model in (36) obeys the following structured counterpart of (as1a).

(**as1**′) *Perturbations in* (36) *are jointly Gaussian, i.e.,* $[\epsilon_A \ \epsilon_y] \sim \mathcal{N}(\mathbf{0}, \mathbf{W}^{-1})$, *as well as independent from* $\mathbf{p}_o^A$ *and* $\mathbf{x}_o$*. Vector* $\mathbf{x}_o$ *has i.i.d. entries with the same prior as in (as1a); and it is independent from* $\mathbf{p}_o^A$, *which has i.i.d. entries drawn from a zero-mean uniform (i.e., non-informative) prior pdf.*

The following optimality claim holds for the WSS-TLS estimator in (28), assured to be equivalent to the solution of problem (26) by Lemma 2.

**Proposition 5:** *(MAP optimality of WSS-TLS). Under (as1*′*) and (as2), the equivalent WSS-TLS problem in* (28) *yields the MAP optimal estimator of* $\mathbf{x}_o$ *and* $\mathbf{p}_A$ *in the structured EIV model* (36).

*Proof:* The proof follows the lines used in proving the MAP optimality of (5) under (as1a) in Proposition 1. The log-prior pdf of $\mathbf{x}_o$ contains an $\ell_1$-norm term as in (8), while the uniform prior on $\mathbf{p}_o^A$ is constant under (as1′). Furthermore, given the structure mapping $[\mathbf{A} \ \mathbf{y}] = \mathbf{S}(\mathbf{p})$, the conditional log-likelihood here can be expressed in terms of $\mathbf{x}$ and $\epsilon^A$, as $\ln p[\mathbf{y}, \mathbf{A}|\mathbf{x}_o = \mathbf{x}, \mathbf{p}_o^A = \mathbf{p}^A + \epsilon^A] = \ln p\left[\epsilon_A = \epsilon^A, \ \epsilon_y = -(\boldsymbol{S}^y)^\dagger[\mathbf{y} - (\mathbf{A} + \sum_{k=1}^{n_A} \epsilon_k^A \mathbf{S}_k^A)\mathbf{x}] = (\boldsymbol{S}^y)^\dagger[\boldsymbol{S}^A(\mathbf{I} \otimes \mathbf{x})\epsilon^A - \mathbf{r}(\mathbf{x})]\right]$. After omitting terms not dependent on $\mathbf{x}$ and $\epsilon^A$, the conditional log-likelihood under the joint Gaussian distribution in (as1′) boils down to half of the quadratic cost in (28). Combining the latter with $\|\mathbf{x}\|_1$ from the log-prior pdf, it follows that maximizing the log-posterior pdf amounts to minimizing the unconstrained sum of the two, which establishes MAP optimality of the WSS-TLS estimator in (28). ∎

## VI. S-TLS APPLICATIONS

In this section, the practical impact of accounting for perturbations present in the data matrix $[\mathbf{A} \ \mathbf{y}]$ will be demonstrated via two sensing applications involving reconstruction of sparse vectors. In both, the perturbation $\mathbf{E}_A$ comes from inaccurate modeling of the underlying actual matrix $\mathbf{A}_o$, while $\mathbf{e}_y$ is due to measurement noise.

### A. Cognitive Radio Sensing

Consider $N_s$ sources located at unknown positions, each transmitting an RF signal with power spectral density (PSD) $\Phi_s(f)$ that is well approximated by a basis expansion model: $\Phi_s(f) = \sum_{\nu=1}^{N_b} x_{s\nu} b_\nu(f)$, where $\{b_\nu(f)\}_{\nu=1}^{N_b}$ are known (e.g., rectangular) basis functions, and $\{x_{s\nu}\}_{s=1}^{N_s}$ are unknown power coefficients. As source positions are also unknown, a Cartesian grid of known points $\{l_g\}_{g=1}^{N_g}$ is adopted to describe *candidate* locations that transmitting radios could be positioned [4], [9]; see also Fig. 1.

The task is to estimate the locations and powers of active sources based on PSD samples measured at $N_r$ cognitive radios (CRs) at known locations $\{\ell_r\}_{r=1}^{N_r}$. Per frequency $f_k$, these samples obey the model

$$\hat{\Phi}_r(f_k) = \sum_{g=1}^{N_g} \gamma_{gr}\Phi_g(f_k) + \sigma_r^2 + e_r(f_k) = [\sum_{g=1}^{N_g}\sum_{\nu=1}^{N_b}\gamma_{gr}b_\nu(f_k)]x_{g\nu} + \sigma_r^2 + e_r(f_k) = \mathbf{a}_r^T(f_k)\mathbf{x}_o + e_r(f_k) \quad (37)$$

where the PSD $\Phi_g(f)$ is nonzero only if a transmitting source is present at $l_g$; $\gamma_{gr}$ represents the channel gain from the candidate source at $l_g$ to the CR at $\ell_r$ that is assumed to follow a known pathloss function of the distance $||l_g - \ell_r||$; $\sigma_r^2$ denotes the known noise variance at receiver $r$; the $N_bN_g \times 1$ vector $\mathbf{a}_r(f_k)$ collects products $\gamma_{gr}b_\nu(f_k)$; vector $\mathbf{x}_o$ contains the $N_bN_g$ unknown power coefficients $x_{g\nu}$; and $e_r(f_k)$ captures the error between the true, $\Phi_r(f_k)$, and estimated, $\hat{\Phi}_r(f_k)$, PSDs.

Estimated PSD samples at $K$ frequencies from all $N_r$ receivers are first compensated by subtracting the corresponding noise variances, and subsequently collected to form the data vector $\mathbf{y}$ of length $m = KN_r$. Noise terms $-e_r(f_k)$ are similarly collected to build the perturbation vector $\boldsymbol{\epsilon}_y$. Likewise, row vectors $\mathbf{a}_r^T(f_k)$ of length $n = N_bN_g$ are concatenated to form the $m \times n$ matrix $\mathbf{A}$. The latter is perturbed (relative to the inaccessible $\mathbf{A}_o$) by a matrix $\mathbf{E}_A$, which accounts for the mismatch between grid location vectors, $\{l_g\}_{g=1}^{N_g}$, and those of the actual sources, $\{\boldsymbol{\kappa}_s\}_{s=1}^{N_s}$. To specify $\mathbf{E}_A$, let $\epsilon_{gr}^s := \gamma_{sr} - \gamma_{gr}$ for the source at $\boldsymbol{\kappa}_s$ closest to $l_g$, and substitute $\gamma_{gr} = \gamma_{sr} - \epsilon_{gr}^s$ into the double sum inside the square brackets of (37). This allows writing $\mathbf{A} = \mathbf{A}_o - \mathbf{E}_A$, where $\mathbf{E}_A$ is affine structured with coefficients $\{\epsilon_{gr}^s\}$ and matrix atoms formed by $\{b_\nu(f_k)\}$. All in all, the setup fits nicely the structured EIV model in (36).

Together with $\{\epsilon_{gr}^s\}$, the support of $\mathbf{x}_o$ estimates the locations of sources, and the nonzero entries of $\mathbf{x}_o$ their transmit powers. Remarkably, this grid-based approach reduces localization – traditionally a nonlinear estimation task – to a linear one, by increasing the problem dimensionality ($N_g \gg N_s$). What is more, $\mathbf{x}_o$ is sparse for two reasons: a) relative to the swath of available bandwidth, the transmitted PSDs are narrowband; hence, the number of nonzero $x_{g\nu}$s is small relative to $N_b$; and (b) the number of actual sources ($N_s$) is much smaller than the number of grid points ($N_g$) that is chosen large enough to localize sources with sufficiently high resolution. Existing sparsity-exploiting approaches to CR sensing rely on BP/Lasso, and do not take into account the mismatch arising due to griding [4], [9]. Simulations in Section VII will demonstrate that sensing accuracy improves considerably if one accounts for grid-induced errors through the EIV model, and compensates for them via the novel WSS-TLS estimators.

## B. DoA Estimation via Sparse Linear Regression

The setup here is the classical one in sensor array processing: plane waves from $N_s$ far-field, narrow-band sources impinge on a uniformly-spaced linear array (ULA) of $N_r$ (possibly uncalibrated) antenna elements. Based on as few $N_r \times 1$ vectors of spatial samples collected across the ULA per time instant $t$ (snapshot), the task is to localize sources by estimating their directions-of-arrival (DoA) denoted as $\{\vartheta_s\}_{s=1}^{N_s}$. High-resolution, (weighted) subspace-based DoA estimators are nonlinear, and rely on the sample covariance matrix of these spatio-temporal samples, which requires a relatively large number of snapshots for reliable estimation especially when the array is not calibrated; see e.g., [21]. This has prompted recent DoA estimators based on sparse linear regression, which rely on a uniform polar grid of $N_g$ points describing *candidate* DoAs $\{\theta_g\}_{g=1}^{N_g}$ [17], [22]. Similar to the CR sensing problem, the $g$-th entry $x_g$ of the $N_g \times 1$ unknown vector of regression coefficients, $\mathbf{x}_{o,t}$, is nonzero and equal to the transmit-source signal power, if a source is impinging at angle $\theta_g$, and zero otherwise.

The $N_r \times 1$ array response vector to a candidate source at DoA $\theta_g$ is $\mathbf{a}(\theta_g) = [1 \ e^{-j\alpha_g} \ \cdots \ e^{-j\alpha_g(N_r-1)}]^T$, where $\alpha_g := 2\pi d \sin(\theta_g)$ denotes the phase shift relative to the source signal wavelength between neighboring ULA elements separated by distance $d$. The per-snapshot received data vector $\mathbf{y}_t$ of length $m = N_r$ obeys the EIV model: $\mathbf{y}_t = (\mathbf{A} + \mathbf{E}_A)\mathbf{x}_{o,t} + (-\mathbf{e}_{y,t})$, where $-\mathbf{e}_{y,t}$ represents the additive noise across the array elements; the $m \times n$ matrix $\mathbf{A} := [\mathbf{a}(\theta_1) \ \cdots \ \mathbf{a}(\theta_{N_g})]$ denotes the grid angle scanning matrix of $n = N_g$ columns; and $\mathbf{E}_A$ represents perturbations arising because DoAs from actual sources do not necessarily lie on the postulated grid points. Matrix $\mathbf{E}_A$ can also account for gain, phase, and position errors of antenna elements when the array is uncalibrated.

To demonstrate how a structured S-TLS approach applies to the DoA estimation problem at hand, consider for simplicity one source from direction $\vartheta_s$, whose nearest grid angle is $\theta_g$; and let $\epsilon_g^s := \vartheta_s - \theta_g$ be the corresponding error that vanishes as the grid density grows large. For small $\epsilon_g^s$, the actual source-array phase shift $\alpha_s := 2\pi d \sin(\theta_g + \epsilon_g^s)$ can be safely approximated as $2\pi d[\sin(\theta_g)\cos(\epsilon_g^s) + \cos(\theta_g)\sin(\epsilon_g^s)] \approx 2\pi d[\sin(\theta_g) + \epsilon_g^s \cos(\theta_g)]$; or, more compactly as $\alpha_s \approx \alpha_g + \epsilon_g \beta_g$, where $\beta_g := 2\pi d \cos(\theta_g)$. As a result, using the approximation $\exp(-j\alpha_s) \approx \exp[-j(\alpha_g + \epsilon_g^s \beta_g)] \approx (1 - j\epsilon_g^s \beta_g) \exp(-j\alpha_g)$, the actual array response vector can be approximated as a linear function of $\epsilon_g^s$; thus, it be expressed as

$$\mathbf{a}(\vartheta_s) = \mathbf{a}(\theta_g) + \epsilon_g^s \boldsymbol{\phi}(\theta_g), \quad \boldsymbol{\phi}(\theta_g) := [0 \ -j\beta_g e^{-j\alpha_g}, \ldots, -j\beta_g(N_r-1)e^{-j\alpha_g(N_r-1)}]^T. \quad (38)$$

With columns obeying (38), the actual array manifold is modeled as $\mathbf{A}_o = \mathbf{A} + \mathbf{E}_A$, where the perturbation matrix is structured as $\mathbf{E}_A = \sum_{g=1}^{N_g} \epsilon_g^s \mathbf{S}_g^A$, with the $N_r \times N_g$ matrix $\mathbf{S}_g^A$ having all zero entries, except for the $g$-th column that equals $\boldsymbol{\phi}(\theta_g)$. With such an array manifold and $\mathbf{S}^y = \mathbf{I}$, the grid-based DoA setup matches precisely the structured EIV model in (36). The simulated tests in the ensuing section will illustrate, among other things, the merits of employing WSS-TLS solvers to estimate $\mathbf{x}_{o,t}$ and $\epsilon_g^s$ based on data collected by possibly antenna arrays. But before this, a final remark is in order.

**Remark 3:** *(Relationships with [16] and [22])* Although $\mathbf{E}_A$ is not explicitly included in the model of existing grid-based approaches, this mismatch has been mitigated either by iteratively refining the grid around the region where sources are present [22], or, by invoking the minimum description length (MDL) test to estimate the number of actual sources $N_s$, followed by spatial interpolation to estimate their DoAs [16]. These remedies require post-processing the initial estimates obtained by sparse linear regression. In contrast, the proposed *structured* S-TLS based approach *jointly* estimates the nonzero support of $\mathbf{x}_{o,t}$ along with grid-induced perturbations. This allows for direct compensation of the angle errors to obtain high-resolution DoA estimates in a single step, and in certain cases without requiring multiple snapshots. Of course, multiple snapshots are expected to improve estimation performance using the matrix S-TLS solver mentioned in Remark 2.

## VII. SIMULATED TESTS

Four simulated tests are presented in this section to illustrate the merits of the S-TLS approach, starting from the algorithms of Section IV.

***Test Case 1:*** *(Optimum vs. suboptimum S-TLS)* The EIV model in (4) is simulated here with a $6 \times 10$ matrix $\mathbf{A}$, whose entries are i.i.d. Gaussian having variance $1/6$, so that the expected $\ell_2$-norm of each column equals 1. The entries of $\mathbf{E}_A$ and $\mathbf{e}_y$ are also i.i.d. Gaussian with variance $0.0025/6$ corresponding to entry-wise signal-to-noise ratio (SNR) of 26dB. Vector $\mathbf{x}_o$ has only nonzero elements in the two first entries: $x_{o,1} = -1.3$ and $x_{o,2} = 5$. Algorithm 1 is tested with $\mu = 5$ against Algorithm 2 implemented with different values of $\lambda$ to obtain a solution satisfying $\|\hat{\mathbf{x}}_{S-TLS}\|_1 = \mu$. For variable tolerance values $\varepsilon$ in Algorithm 1-b, the attained minimum cost $f(\mathbf{x})$ in (11) is plotted in Fig. 2. To serve as a benchmark, a genie-aided globally optimum scheme is also tested with the support of $\mathbf{x}$ known and equal to that of $\mathbf{x}_o$. Specifically, the genie-aided scheme minimizes $f(\mathbf{x})$ over all points with $\ell_1$-norm equal to $\mu$, and all entries being 0 except for the first two. Using the equivalence between (11) and (12), the genie-aided scheme per iteration amounts to minimizing a scalar quadratic program under linear constrains, which is solved efficiently using the interior-point optimization routine in [27].

Fig. 2 shows that as $\varepsilon$ becomes smaller, the minimum achieved value $f(\mathbf{x})$ decreases monotonically, and drops sharply to the global minimum attained by the genie-aided bisection scheme. Interestingly, the alternating descent algorithm that guarantees convergence to a stationary point, exhibits performance comparable to the global algorithm. For this reason, only the alternating descent algorithm is used in all subsequent tests. Next, S-TLS estimates are compared with those obtained via BP/Lasso and (regularized) TLS in the context of the CR sensing and array processing applications outlined in Section VI.

*Test Case 2: (S-TLS vs. Lasso vs. TLS)* The setup here is also based on the EIV model (4), with $\mathbf{A}$ of size $20 \times 40$ having i.i.d. Gaussian entries; and $\mathbf{x}_o$ having 5 nonzero i.i.d. standardized Gaussian entries. By averaging results over 200 Monte Carlo runs, the S-TLS solution is compared against the Lasso one for 20 values of $\lambda$ (uniformly spaced in log-scale), based on the $\ell_2$, $\ell_1$, and $\ell_0$ errors of the estimated vectors relative to $\mathbf{x}_o$. (The $\ell_0$ error equals the percentage of entries for which the support of the two vectors is different.) Fig. 3 corroborates the improvement of S-TLS over Lasso, especially in the $\ell_0$ norm. Fig. 3(c) further demonstrates that over a range of moderate $\lambda$ values, S-TLS consistently outperforms Lasso in recovering the true support of $\mathbf{x}_o$. For high $\lambda$'s, both estimates come close to the all-zero vector, so that the $\ell_0$ errors become approximately the same, even though the $\ell_2$ and $\ell_1$ errors are smaller for Lasso. However, for both error norms S-TLS has a slight edge over moderate values of $\lambda$.

Receiver operating characteristic (ROC) curves are plotted in Fig. 3(d) to illustrate the merits of S-TLS and Lasso over (regularized) TLS in recovering the correct support. The "best" $\lambda$ for the S-TLS and Lasso algorithms is chosen using cross-validation [25]. As TLS cannot be applied to under-determined systems, a $40 \times 40$ matrix $\mathbf{A}$ is selected. Since TLS and LS under an $\ell_2$-norm constraint $\|\mathbf{x}\|_2 \leq \mu$ are known to be equivalent when $\mu$ is small [26], the regularized TLS is tested using the function `lsqi` for regularized LS from [18]. The probability of correct detection, $P_d$, is calculated as the probability of identifying correctly the support over nonzero entries of $\mathbf{x}_o$, and the probability of false alarms, $P_{fa}$, as that of incorrectly deciding zero entries to be nonzero. The ROC curves in Fig. 3(d) demonstrate the advantage of Lasso, and more clearly that of S-TLS, in recovering the correct support.

*Test Case 3: (CR Spectrum Sensing)* This simulation is performed with reference to the CR network in the region $[0\ 1] \times [0\ 1]$ in Fig. 1. The setup includes $N_r = 4$ CRs deployed to estimate the power and location of a single source with position vector $[0.4\ 0.6]$, located at the center of four neighboring grid points. The CRs scan $K = 128$ frequencies from 15MHz to 30MHz, and adopt the basis expansion model in Section VI-A with $N_b = 16$ rectangular $b_\nu(f)$ functions, each of bandwidth 1MHz. The actual source only transmits over the $\nu = 6$-th band. The channel gains are exponentially decaying in distances with exponent $-1/2$. The received data are generated using the transmit PSD described earlier, a regular Rayleigh fading channel with 6 taps, and additive white Gaussian receiver noise at SNR=0dB. Receive-PSDs are obtained using exponentially weighted periodograms (with weight 0.99) averaged over 1,000 coherence blocks; see also [4] for more details of a related simulation. The WSS-TLS approach is used to account for perturbations $\epsilon_{gr}^s$ in the channel gains. A diagonal matrix $\mathbf{W}$ is used with each diagonal entry equal to $\hat{\sigma}_\epsilon^{-2}$ (inversely proportional to the average of of sample variances of $\epsilon_{gr}^s$).

With $\lambda$ chosen as in [11], both Lasso and WSS-TLS identify the active frequency band correctly (only the entries $\{x_{g6}\}_{g=1}^{16}$ were estimated as nonzero). However, Lasso identifies four transmitting sources at positions $[0.3(0.5)\ 0.5(0.7)]$, the four grid points closest to $[0.4\ 0.6]$. WSS-TLS returns only one source at position $[0.5\ 0.5]$, along with the estimated $\hat{\epsilon}_{gr}^s$ that yields $\hat{\gamma}_{sr} = \hat{\epsilon}_{gr}^s + \gamma_{gr}$. Concatenate the latter to form $\hat{\gamma}_s$ of length $N_r = 4 \ll m$. Using a refined grid of 25 points uniformly spaced over the "zoom-in" region $[0.3\ 0.7] \times [0.3\ 0.7]$ centered at $[0.5\ 0.5]$, correlation coefficients between $\hat{\gamma}_s$ and those of each candidate point are evaluated. The source position is estimated as the point with maximum correlation coefficient, which for WSS-TLS occurs at the true location $[0.4\ 0.6]$. To illustrate graphically the two alternatives, the estimated maps of the spatial PSDs at the 6th frequency band are plotted in Fig. 4(a) using the Lasso, and in Fig. 4(b) using WSS-TLS. The marked point indicates the actual source location $[0.4\ 0.6]$ in both maps. Unlike Lasso, the WSS-TLS identifies correctly the true position of the source.

***Test Case 4:*** *(DoA Estimation)* The setup here entails a ULA consisting of $N_r = 8$ antenna elements with inter-element spacing $d = 1/2$, and a grid of $N_g = 90$ scanning angles from $-90°$ to $90°$ wrt the array boresight. Two sources ($N_s = 2$) of unit amplitude impinge from angles $\vartheta_s = 1°$ and $-9°$, both $1°$ off their nearest grid DoAs. As in the single-snapshot test in [22], the SNR is set to 20dB. The variance of $\epsilon_g^s$ in (38) is obtained from the uniform distribution in $[-1°, 1°]$. Selecting $\lambda$ according to the noise level as in [22], Lasso returns four nonzero entries, two around each source at $\vartheta_s \pm 1°$; while WSS-TLS gives two nonzero $\theta_g$ estimates at $-10°$ ($g = 40$) and $0°$ ($g = 45$), along with perturbation estimates $\hat{\epsilon}_{40}^s$ and $\hat{\epsilon}_{45}^s$. Using the latter, the DoAs are estimated as $\hat{\vartheta}_s := \hat{\theta}_g + \hat{\epsilon}_g^s$ for $g = 40,\ 45$. The angle spectra using Lasso, and WSS-TLS with estimated $\hat{\vartheta}_s$, are compared in Fig. 5(a). The two black arrows depict the actual source angles, and benchmark the true angular spectrum.

To further illustrate the merits of WSS-TLS in estimating correctly the closest grid point and subsequently each DoA, the sample variance of a DoA estimate is plotted versus SNR in Fig. 5(b) using Monte Carlo runs, each with a single source randomly placed over $[-1°, 1°]$. Both WSS-TLS and Lasso are post-processed by interpolating peaks in the obtained spectra from two nearest grid points, linearly weighted by the estimated amplitudes as in [17]. Both curves confirm that WSS-TLS outperforms the Lasso. More interestingly, the two WSS-TLS curves almost coincide, which further corroborates that WSS-TLS manages in a single step to identify correctly the support of $\mathbf{x}_{o,t}$ without requiring post processing.

## VIII. CONCLUDING REMARKS

An innovative approach was developed in this paper to account for sparsity in estimating coefficient vectors of fully-perturbed linear regression models. This approach enriches TLS criteria that have been traditionally used to fit such models with the ability to handle under-determined linear systems. The novel S-TLS framework also enables sparsity-exploiting approaches (CS, BP, and Lasso) to cope with perturbations present not only in the data but also in the regression matrix.

Near-optimum and reduced-complexity suboptimum solvers with global and local convergence guarantees were also developed to optimize the generally nonconvex S-TLS criteria. They rely on bisection, branch-and-bound, or coordinate descent iterations, and have universal applicability regardless of whether perturbations are modeled as deterministic or random. Valuable generalizations were also provided when prior information is available on the deterministic structure or statistics of the associated (augmented) data matrix. Under specific statistical models with errors-in-variables, the resultant (generally weighted and structured) S-TLS estimators were proved to be optimal in the MAP sense. Simulated tests corroborated the analytical claims, compared competing alternatives, and demonstrated the practical impact of the novel S-TLS framework to grid-based sparsity-exploiting approaches for cognitive radio sensing, and direction-of-arrival estimation with possibly uncalibrated antenna arrays.

Interesting topics to explore in future research, include performance analysis for the proposed S-TLS algorithms, and online implementations for S-TLS optimal adaptive processing.

**Algorithm 1-a** (BB): Input $\mathbf{y}$, $\mathbf{A}$, $a$, and $\delta$. Output a $\delta$-optimal solution $\mathbf{x}_g^\star$ of (12)

Set $\mathbf{x}_L = -\mu\mathbf{1}$, $\mathbf{x}_U = \mu\mathbf{1}$, $\Omega := \{(\mathbf{x}_L, \mathbf{x}_U, -\infty)\}$, and initialize with $\mathcal{U} = \infty$.
**repeat**

    Let $(\mathbf{x}_L, \mathbf{x}_U, c)$ be one triplet of $\Omega$ with the smallest $c$; and set $\Omega = \Omega \backslash (\mathbf{x}_L, \mathbf{x}_U, c)$.
    Solve (13) locally to obtain $\hat{\mathbf{x}}_g^\star$.
    **if** $g(\hat{\mathbf{x}}_g^\star, a) < \mathcal{U}$ **then**

        Set $\mathcal{U} = g(\hat{\mathbf{x}}_g^\star, a)$ and $\mathbf{x}_g^\star = \hat{\mathbf{x}}_g^\star$. {update the minimum}
    **end if**
    Minimize globally the convex $g_L(\mathbf{x}, a)$ in (14) with the optimum $\mathbf{D}$ in (15), to obtain $\check{\mathbf{x}}_g^\star$ and $\mathcal{L} := g_L(\check{\mathbf{x}}_g^\star, a)$.

    **if** $\mathcal{U} - \mathcal{L} > \delta$ {need to split} **then**

        Find $i = \arg\max_n ([\mathbf{x}_U]_n - [\mathbf{x}_L]_n)$.
        Set $\mathbf{x}_{L,1}$ ($\mathbf{x}_{U,1}$) and $\mathbf{x}_{L,2}$ ($\mathbf{x}_{U,2}$) equal to $\mathbf{x}_L$ ($\mathbf{x}_U$) except for the $i$-th entry. {split the maximum separation}

        Set $[\mathbf{x}_{L,1}]_i = [\mathbf{x}_L]_i$, $[\mathbf{x}_{U,1}]_i = ([\mathbf{x}_U]_i - [\mathbf{x}_L]_i)/2$, $[\mathbf{x}_{L,2}]_i = ([\mathbf{x}_U]_i - [\mathbf{x}_L]_i)/2$, and $[\mathbf{x}_{U,2}]_i = [\mathbf{x}_U]_i$.
        Augment the set of unsolved boxes $\Omega = \Omega \bigcup \{(\mathbf{x}_{L,1}, \mathbf{x}_{U,1}, \mathcal{L}), (\mathbf{x}_{L,2}, \mathbf{x}_{U,2}, \mathcal{L})\}$.
    **end if**
**until** $\Omega = \emptyset$

**Algorithm 1-b** (Bisection): Input $\mathbf{y}$, $\mathbf{A}$, and tolerances $\varepsilon$ and $\delta$. Output an $\varepsilon$-optimal solution $\mathbf{x}_\varepsilon^\star$ to (10)

1: Set $l_0 = 0$, $u_0 = \|\mathbf{y}\|_2^2$, iteration index $i = 0$, and initialize the achievable cost $f_m = u_0$ with $\mathbf{x}_\varepsilon^\star = \mathbf{0}_{n \times 1}$.
2: **repeat**
3:     Let $a = (l_i + u_i)/2$ and call Algorithm 1-a to find a feasible $\delta$-optimal solution $\mathbf{x}_g^\star$ to (12).
4:     Calculate $f_g = f(\mathbf{x}_g^\star)$, and update the iteration $i = i + 1$.
5:     **if** $f_g < f_m$ **then**
6:         Set $f_m = f_g$ and $\mathbf{x}_\varepsilon^\star = \mathbf{x}_g^\star$. {update the minimum}
7:     **end if**
8:     **if** $g(\mathbf{x}_g^\star, a) \leq 0$ **then**
9:         Update $u_i = a$ and $l_i = l_{i-1}$.
10:     **else if** $g(\mathbf{x}_g^\star, a) \geq \delta$ **then**
11:         Update $l_i = a$ and $u_i = u_{i-1}$.
12:     **else**
13:         Update $l_i = a - \delta$ and $u_i = u_{i-1}$.
14:     **end if**
15:     Set $u_i = \min(u_i, f_g)$.
16: **until** $u_i - l_i \leq \varepsilon$

**Algorithm 2** (CD): Input $\mathbf{y}$, $\mathbf{A}$, and coefficient $\lambda$. Output the iterates $\mathbf{E}(i)$ and $\mathbf{x}(i)$ upon convergence.

Initialize with $\mathbf{E}(0) = \mathbf{0}_{m \times n}$ and $\mathbf{x}(-1) = \mathbf{0}_{n \times 1}$
**for** $i = 0, 1, \ldots$ **do**
   **for** $\nu = 1, \ldots, n$ **do**
     Compute the residual $\mathbf{e}_\nu(i)$ as in (20).
     Update the scalar $x_\nu(i)$ via (21).
   **end for**
   Update the iterate $\mathbf{E}(i+1)$ as in (19).
**end for**

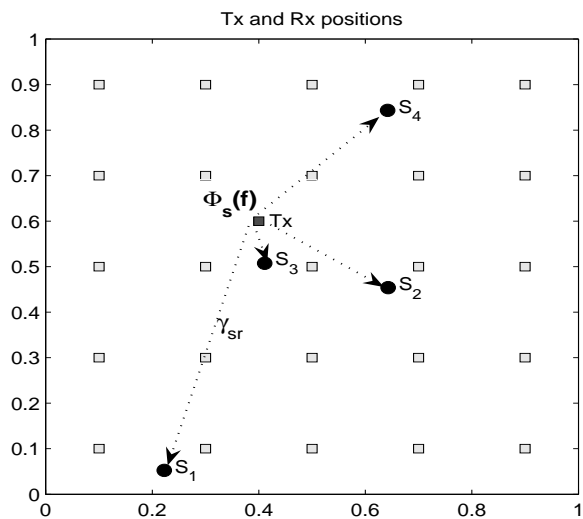

Fig. 1. Grid topology with $N_g = 25$ candidate locations, $N_s = 1$ transmitting source, and $N_r = 4$ receiving CRs.

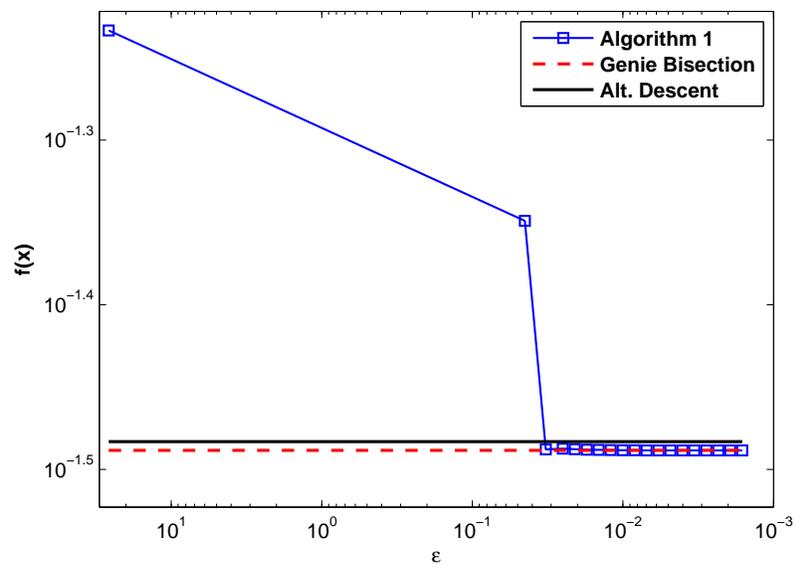

Fig. 2. Attained $f(\mathbf{x})$ for variable tolerance values $\varepsilon$ by the global Algorithm 1-b, compared to the alternating descent local algorithm, and the genie-aided global solver.

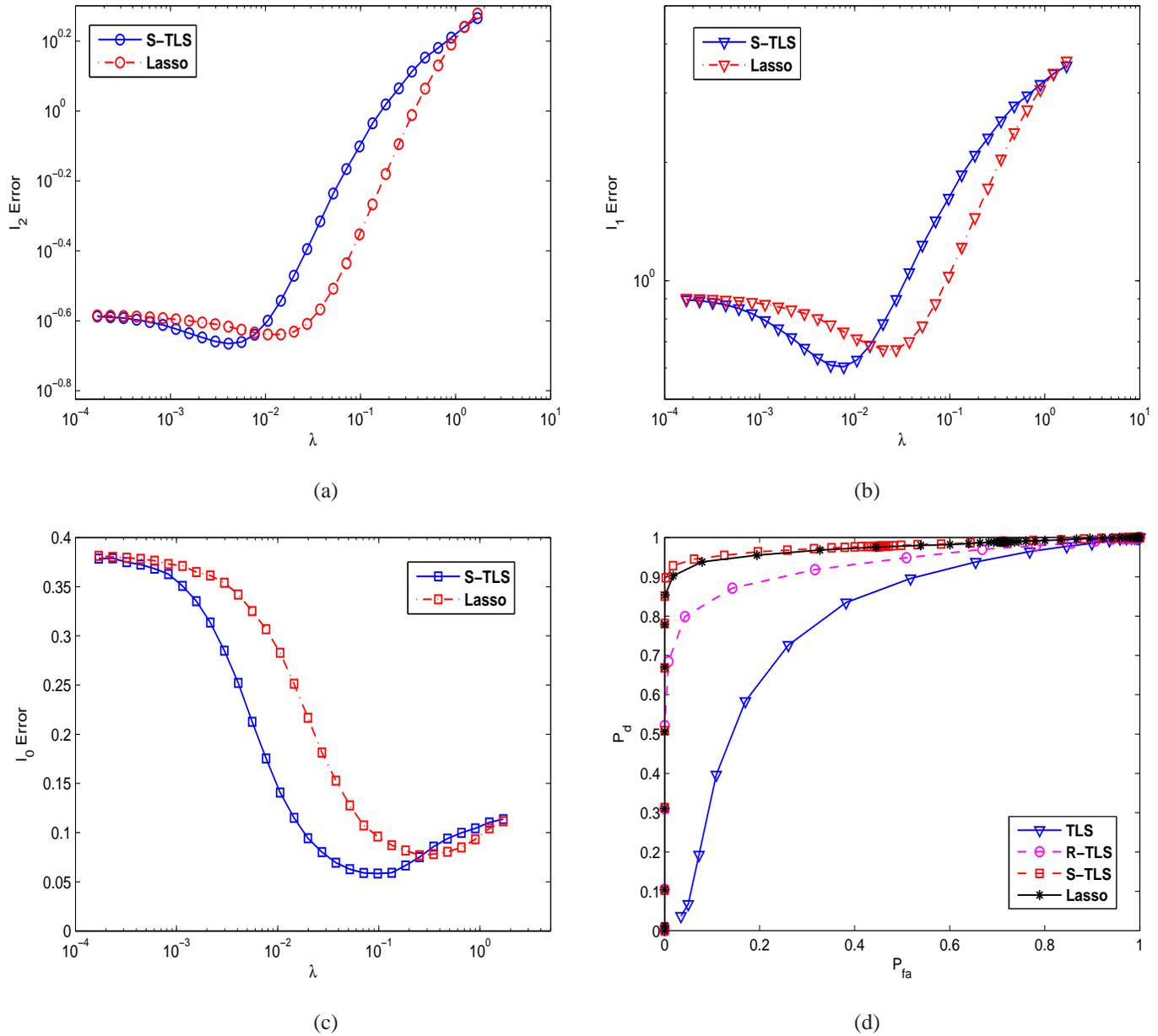

Fig. 3. Comparison between S-TLS and Lasso in terms of: (a) $\ell_2$-norm, (b) $\ell_1$-norm, and (c) $\ell_0$-norm of the estimation errors; (d) Probability of detection versus probability of false alarms for the TLS, $\ell_2$ regularized (R-)TLS, S-TLS and Lasso algorithms.

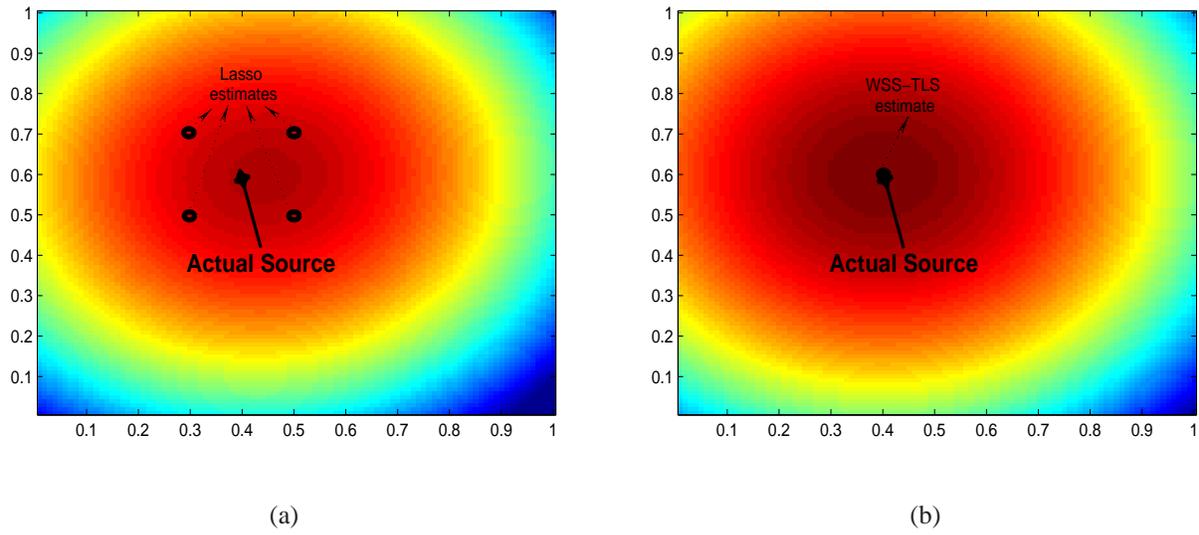

Fig. 4. Comparison between PSD maps estimated by (a) Lasso, and (b) WSS-TLS for the CR network in Fig. 1.

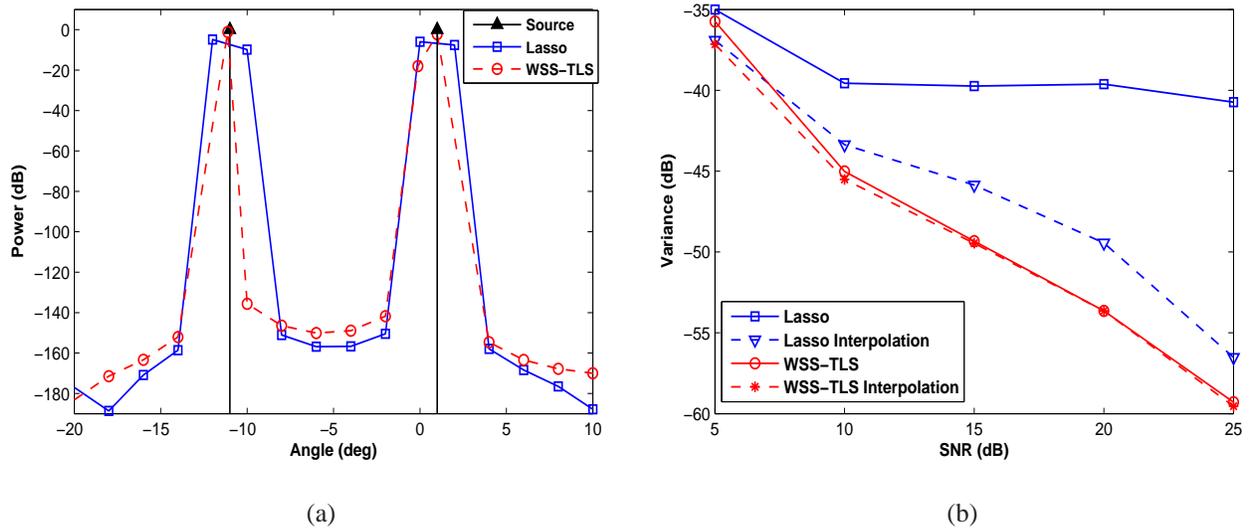

Fig. 5. (a) Angular spectra estimated using Lasso and WSS-TLS as compared to the actual transmission pattern; (b) Comparison of angle estimation variances of Lasso, WSS-TLS, without and with interpolation.